\documentclass[onecolumn]{IEEEtran}

\usepackage{graphicx}
\usepackage{tikz}
 \usetikzlibrary{positioning}
 \usetikzlibrary{calc}
 \usetikzlibrary{shapes}
 \usetikzlibrary{patterns}
\usepackage{pgfplots}
 \pgfplotsset{compat=1.14}
 \usepgfplotslibrary{fillbetween}
\usepackage{dsfont}
\usepackage{adjustbox}
\usepackage{enumitem}
\usepackage{amsmath,amssymb,amsthm}
\usepackage{bbm}
\usepackage{standalone}
\usepackage{multirow}
\usepackage[linesnumbered,ruled]{algorithm2e}
\setenumerate{parsep=6pt,itemsep=0pt,leftmargin=*,labelsep=5.5mm} 
\setlist[description]{itemsep=0mm}   

\usepackage{hyperref}
\hypersetup{colorlinks=true,urlcolor=blue,citecolor=red,linkcolor=blue}

\usepackage{amsmath,amssymb,amsthm}
    \theoremstyle{definition}
	\newtheorem{lemma}{Lemma}
	\newtheorem{theorem}{Theorem}
	
	\newtheorem{example}{Example}
    \newtheorem{remark}{Remark}
    \newtheorem{definition}{Definition}

\def\mm{m}
\def\uboldb{\underline{\boldsymbol{b}}}
\def\uboldB{\underline{\boldsymbol{B}}}
\def\uboldbt{\tilde{\underline{\boldsymbol{b}}}}

\def\ubtt{\tilde{\underline{b}}}

\def\define{\stackrel{\Delta}{=}}

\def\boldb{\boldsymbol{b}}
\def\boldB{\boldsymbol{B}}
\def\boldc{\boldsymbol{c}}
\def\boldC{\boldsymbol{C}}

\def\boldx{\boldsymbol{x}}
\def\boldX{\boldsymbol{X}}

\def\calA{\mathcal{A}}
\def\calB{\mathcal{B}}

\def\calS{\mathcal{S}}

\def\calX{\mathcal{X}}
\def\calY{\mathcal{Y}}

\def\snr{\text{SNR}}

\def\Rc{R_{\text{c}}}

\def\gmi{\text{GMI}}
\def\lm{\text{LM}}
\def\rbmd{R_\text{BMD}}

\def\usep{\tilde{\underline{s}}'}
\def\BepsV{\mathcal{B}_{V,\varepsilon}^n}
\def\BepsSY{\mathcal{B}_{SY,\varepsilon}^n}

\newcommand{\Aeps}{\mathcal{A}^{n}_{\varepsilon}}
\newcommand{\Beps}{\mathcal{B}^{n}_{\varepsilon}}
\newcommand{\ua}{\underline{a}}
\newcommand{\uA}{\underline{A}}
\newcommand{\uap}{\underline{\tilde{a}}}
\newcommand{\us}{\underline{s}}
\newcommand{\usp}{\underline{\tilde{s}}}
\newcommand{\use}{\underline{s}'}
\newcommand{\uSe}{\underline{S}'}
\newcommand{\ust}{\underline{s}''}
\newcommand{\uSt}{\underline{S}''}
\newcommand{\ustp}{\underline{\tilde{s}}''}

\newcommand{\ux}{\underline{x}}
\newcommand{\uxp}{\underline{\tilde{x}}}
\newcommand{\uy}{\underline{y}}
\newcommand{\uY}{\underline{Y}}

\newcommand{\ube}{\underline{b}_1}
\newcommand{\ubep}{\underline{\tilde{b}}_1}

\newcommand{\ubt}{\underline{b}_2}
\newcommand{\ubtp}{\underline{\tilde{b}}_2}

\newcommand{\Pe}{P_e}
\newcommand{\oPe}{\overline{P}_e}
\newcommand{\SNR}{\mbox{SNR}}

\newcommand{\uS}{\underline{S}}
\newcommand{\uu}{\underline{u}}
\newcommand{\ub}{\underline{b}}
\newcommand{\uB}{\underline{B}}
\newcommand{\uU}{\underline{U}}
\newcommand{\uv}{\underline{v}}
\newcommand{\uV}{\underline{V}}

\newcommand{\uX}{\underline{X}}

\tikzstyle{point-visible} = [debugpoint,inner sep=0pt, minimum size = 2,fill=red] 
\tikzstyle{point-invisible} = [coordinate]
\tikzstyle{point} = [point-invisible] 

\begin{document}

\title{Achievable Information Rates for Probabilistic Amplitude Shaping: An Alternative Approach via Random Sign-Coding Arguments} 

\author{\IEEEauthorblockN{Yunus Can G\"{u}ltekin,
Alex Alvarado, and 
Frans M. J. Willems}\\
\IEEEauthorblockA{Information and Communication Theory Laboratory (ICTLab)}\\
\IEEEauthorblockA{Signal Processing Systems (SPS) Group, Department of Electrical Engineering}\\
\IEEEauthorblockA{Eindhoven University of Technology, The Netherlands}\\
\IEEEauthorblockA{Emails: \{y.c.g.gultekin, a.alvarado, f.m.j.willems\}@tue.nl }
\thanks{{\bf Acknowledgements:} The work of Y.C. G\"{u}ltekin and A. Alvarado has received funding from the European Research Council (ERC) under the European Union's Horizon 2020 research and innovation programme (grant agreement No 757791).}
}

\markboth{Draft}{}

\maketitle

\begin{abstract}
Probabilistic amplitude shaping (PAS) is a coded modulation strategy in which constellation shaping and channel coding are combined. 
PAS has attracted considerable attention in both wireless and optical communications.
Achievable information rates (AIRs) of PAS have been investigated in the literature using Gallager's error exponent approach.
In particular, it has been shown that PAS achieves the capacity of the additive white Gaussian noise channel (B\"{o}cherer, 2018). 
In this work, we~revisit the capacity-achieving property of PAS and derive AIRs using weak typicality.
Our objective is to provide alternative proofs based on random sign-coding arguments that are as constructive as possible. 
Accordingly, in our proofs, only some signs of the channel inputs are drawn from a random code, while the remaining signs and amplitudes are produced constructively.
We~consider both symbol-metric and bit-metric decoding.
\end{abstract}


\section{Introduction}
Coded modulation (CM) refers to the design of forward error correction (FEC) codes and high-order modulation formats, which are combined to reliably transmit more than one bit per channel use.
Examples of CM strategies include multilevel coding (MLC)~\cite{Imai1977_TransIT_multilevel,Wachsmann1999_MLcodes} in which each address bit of the signal point is protected by an individual binary FEC code, and~trellis CM~\cite{Ungerbock1982_TransIT_TCM}, which combines the functions of a trellis-based channel code and a modulator.
Among many CM strategies, bit-interleaved CM (BICM)~\cite{Zehavi1992_TCOM_bicm,Caire1998_BICM}, which combines a high-order modulation format with a binary FEC code using a binary labeling strategy and uses bit-metric decoding (BMD) at the receiver, is the de-facto standard for CM.
BICM is included in multiple wireless communication standards such as the IEEE 802.11~\cite{IEEE80211_2016} and the DVB-S2
~\cite{DVBS2}.
BICM is also currently the de-facto CM alternative for fiber optical~communications.

Proposed in~\cite{BochererSS2015_ProbAmpShap}, probabilistic amplitude shaping (PAS) integrates constellation shaping into existing BICM systems.
The shaping gap that exists for the additive white Gaussian noise (AWGN) channel~\cite[Ch. 9]{CoverT2006_ElementsofInfoTheo} can be closed with PAS.
To this end, an~amplitude shaping block converts binary information strings into shaped amplitude sequences in an invertible manner.
Then, a~systematic FEC code produces parity bits encoding the binary labels of these amplitudes.
These parity bits are used to select the signs, and~the combination of the amplitudes and the signs, i.e.,~probabilistically shaped channel inputs, are transmitted over the channel.
PAS has attracted considerable attention in fiber optical communications due to its availability of providing rate adaptivity~\cite{Buchali2016_RateAdaptReachIncrease,idler2017_fieldtrialPS}.

Achievable information rates (AIRs) of PAS have been investigated in the literature~\cite{Bocherer2018_AchvRates4ProbShap,Bocherer2018_PrinciplesCM,Amjad2018_InfRatesErrExponents}.
It has been shown that the capacity of the AWGN channel can be achieved with PAS, e.g.,~in~\cite[Example~10.4]{Bocherer2018_PrinciplesCM}.
The achievability proofs in the literature are based on Gallager's error exponent approach~\cite[Ch. 5]{Gallager1968_ITbook} or on strong typicality~\cite[Ch. 1]{Kramer2008_TopicsMUInfoTheo}.

In this work, we provide a {random sign-coding} framework based on weak-typicality that contains the achievability proofs relevant for the PAS architecture.
We also revisit the capacity-achieving property of PAS for the AWGN channel.
As explained in Section~\ref{ssec:contribution}, the~first main contribution of this paper is to provide a framework that combines the constructive approach to amplitude shaping with randomly-chosen error-correcting codes, where the randomness is concentrated only in the choice of the signs.
The second contribution is to provide a unifying framework of achievability proofs to bring together PAS results that are somewhat scattered in the literature, using a single proof technique, which we call the random sign-coding~arguments.

This work is organized as follows.
In Section~\ref{sec:relatedwork}, we briefly summarize the related literature on CM, AIRs, and~PAS and state our contribution.
In Section~\ref{sec:prelim}, we provide some background information on typical sequences and define a modified (weakly) typical set. 
In Section~\ref{sec:randomsigncoding}, we explain the random sign-coding setup.
Finally in Section~\ref{sec:airsofsigncoding}, we provide random sign-coding arguments to derive AIRs for PAS and, consequently, show that it achieves the capacity of a discrete-input memoryless channel with a symmetric capacity-achieving distribution. 
Conclusions are drawn in Section~\ref{sec:conc}.
 
\section{Related Work and Our~Contribution}\label{sec:relatedwork}
\vspace{-6pt}
\subsection{Notation}
Capital letters $X$ are used to denote random variables, while lower case letters $x$ are used to denote their realizations.
Underlined capital and lower case letters $\uX$ and $\ux$ are used to denote random vectors and their realizations, respectively.
Boldface capital and lower case letters $\boldX$ and $\boldx$ are used to denote collections of random variables and their realizations, respectively.
Underlined boldface capital and lower case letters $\underline{\boldX}$ and $\underline{\boldx}$ are used to denote collections of random vectors and their realizations, respectively.
Element-wise multiplication of $\ux$ and $\uy$ is denoted by $\ux \otimes \uy$.
Calligraphic~letters $\calX$ represent sets, while $\calX \calY = \{xy: x\in\calX, y \in \calY\}$.
We denote by $\calX^n$ the $n$-fold Cartesian product of $\calX$ with itself, while $\calX \times \calY$ is the Cartesian product of $\calX$ and $\calY$.
Probability density and mass functions over $\calX$ are denoted by $p(x)$.
We use $\mathds{1}[\cdot]$ to indicate the indicator function, which is one when its argument is true and zero otherwise.
The entropy of $X$ is denoted by $H(X)$ (in bits), the~expected value of $X$ by $E[X]$.

\subsection{Achievable Information~Rates}\label{sec:AIRs}
For a memoryless channel that is characterized by an input alphabet $\calX$, input distribution $p(x)$, and~channel law $p(y|x)$, the~maximum AIR is the mutual information (MI) $I(X;Y)$ of the channel input $X$ and output $Y$.
Consequently, the~capacity of this channel is defined as $I(X;Y)$ maximized over all possible input distributions $p(x)$, typically under an average power constraint, e.g.,~in~\cite[Sec. 9.1]{CoverT2006_ElementsofInfoTheo}.  
The MI can be achieved, e.g.,~with MLC and multi-stage decoding~\cite{Imai1977_TransIT_multilevel,Wachsmann1999_MLcodes}.

In BICM systems, channel inputs are uniquely labeled with $\log_2|\calX|=(m+1)$-bit binary strings.
Here, we assume that $|\calX|$ is an integer power of two.
At the transmitter, the~output of a binary FEC code is mapped to channel inputs using this labeling strategy.
At the receiver, BMD is employed, i.e.,~binary~labels $\boldC = (C_1,C_2,\cdots, C_{m+1})$ are assumed to be independent, and~consequently, the~symbol-wise decoding metric is written as the product of bit-metrics:
\begin{equation}
 \mathbbm{q}(x,y) = \prod_{i=1}^{m+1} \mathbbm{q}_i(c_i,y). \label{eq:prodmetric}
\end{equation}
Since the metric in \eqref{eq:prodmetric} is in general not proportional to $p(y|x)$, i.e.,~there is a mismatch between the actual channel law and the one assumed at the receiver, this setup is called {mismatched decoding}.

Different AIRs have been derived for this so-called mismatched decoding setup.
One of these is the generalized MI (GMI)~\cite{Kaplan1993_RatesExponentsCompondChan,Merhav1994_InfoRatesMismDecod}:
\begin{equation}
 \gmi \left(p(x)\right) = \max_{s\geq0} E\left[ \log \frac{\left[ \mathbbm{q}(X,Y)\right]^s}{\sum_{x\in\calX} p(x) \left[ \mathbbm{q}(x,Y)\right]^s}\right], \label{eq:GMI}
\end{equation}
which reduces to~\cite[Thm. 4.11, Coroll. 4.12]{Szczecinski2015_BICMbook} and~\cite{Martinez2009_TransIT_bicm}:
\begin{equation}
 \text{GMI}\left( p(c_1)p(c_2)\cdots p(c_{m+1}) \right) = \sum_{i=1}^{m+1} I(C_i;Y) \label{eq:independentgmi}
\end{equation}
when the bit levels are independent at the transmitter, i.e.,~$p(x)=p(\boldc)=p(c_1)p(c_2)\cdots p(c_{m+1})$ where $\boldc = (c_1, c_2, \cdots, c_{m+1})$, and:
\begin{equation}
 \mathbbm{q}_i(c_i,y) = p(y|c_i). \label{eq:gmibitmetrics}
\end{equation}
The rate \eqref{eq:independentgmi} is achievable for both uniform and shaped bit levels~\cite{Caire1998_BICM,FabregasM2010_BICMwShaping}.
The problem of computing the bit level distributions that maximize the GMI in \eqref{eq:independentgmi} was shown to be nonconvex in~\cite{AlvaradoBA2011_HighSNRforBICMcapacity}.
The parameter that maximizes \eqref{eq:GMI} to obtain \eqref{eq:independentgmi} is $s=1$.

Another AIR for mismatched decoding is the LM (lower bound on the mismatch capacity) rate~\cite{Merhav1994_InfoRatesMismDecod,Peng2012PhdThesis}:
\begin{equation}
 \lm \left(p(x)\right) = \max_{s\geq0,r(\cdot)} E\left[ \log \frac{\left[ \mathbbm{q}(X,Y)\right]^s r\left(X\right)}{\sum_{x\in\calX} p(x) \left[ \mathbbm{q}(x,Y)\right]^s r\left(x\right)}\right], \label{eq:LMrate}
\end{equation}
where $r(\cdot)$ is a real-valued cost function defined on $\calX$.
The expectations in \eqref{eq:GMI} and \eqref{eq:LMrate} are taken with respect to $p(x,y)$.

When there is dependence among bit levels, i.e.,~$p(x) = p(\boldc) \neq p(c_1)p(c_2)\cdots p(c_{m+1})$, the~rate~\cite{Bocherer2014_ProbSigShapForBMD,Bocherer2014_ProbSigShapForBMD_Arxiv_mdpi}:
\begin{equation}
\rbmd\left(p(x)\right) = H\left(\boldC \right) - \sum_{i=1}^{m+1} H(C_i|Y) \label{eq:rbmd}
\end{equation}
has been shown to be achievable by BMD for any joint input distribution $p(\boldc) = p(c_1,c_2,\cdots,c_{m+1})$.
In~\cite{Bocherer2014_ProbSigShapForBMD,Bocherer2014_ProbSigShapForBMD_Arxiv_mdpi}, the~achievability of \eqref{eq:rbmd} was derived using random coding arguments based on strong typicality~\cite[Ch. 1]{Kramer2008_TopicsMUInfoTheo}.
Later in~\cite[Lemma 1]{Bocherer2014_AchRatesForShapedBMD_mdpi}, it was shown that \eqref{eq:rbmd} is an instance of the so-called LM rate \eqref{eq:LMrate} for $s=1$, the~symbol decoding metric \eqref{eq:prodmetric}, bit decoding metrics \eqref{eq:gmibitmetrics}, and~the cost function:
\begin{equation}
 r(c_1,c_2,\cdots,c_{m+1}) = \frac{\prod_{i=1}^{m+1} p(c_i) } {p(c_1,c_2,\cdots,c_{m+1})}.
\end{equation}
We note here that $\rbmd$ in \eqref{eq:rbmd} can be negative as discussed in~\cite[Sec. II-B]{Bocherer2014_AchRatesForShapedBMD_mdpi}. 
In such cases, $\rbmd$~cannot be considered as an achievable rate.
To avoid this, $\rbmd$ is defined as the maximum of~\eqref{eq:rbmd} and zero in~\cite[Eq. (1)]{Bocherer2014_AchRatesForShapedBMD_mdpi}.

\subsection{Probabilistic Amplitude Shaping: Model}
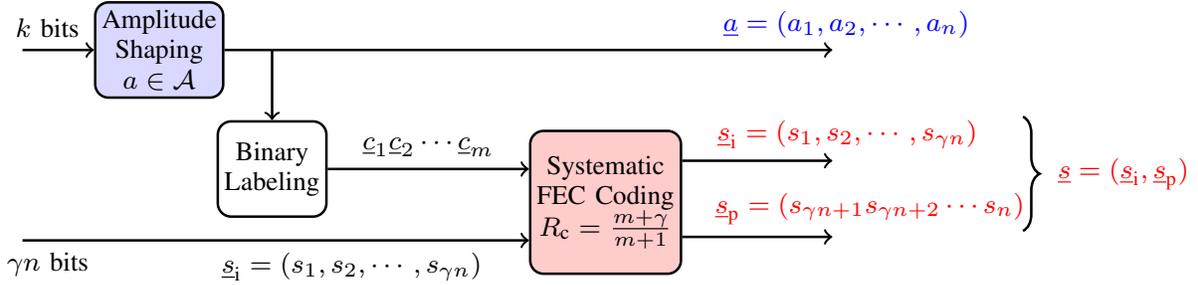
\begin{figure}[t]
\centering
\resizebox{0.9\columnwidth}{!}{
\begin{tikzpicture}[
line width=0.75pt,
block/.style={rectangle, rounded corners,draw,inner sep=2pt,minimum width=10mm, minimum height=10mm,font=\footnotesize,align=center},
block_circ/.style={ellipse, rounded corners,draw,inner sep=2pt,minimum width=5mm, minimum height=5mm,align=center},
block_ell/.style={ellipse, rounded corners,draw,inner sep=2pt,minimum width=10mm, minimum height=7mm,font=\footnotesize,align=center},
line_arrow/.style = {draw,->,font=\footnotesize},
line/.style = {draw,-,font=\footnotesize}
]
\pgfmathsetmacro{\nodedist}{1}
\coordinate (start) at (0,0);
\coordinate (start_below) at (0,-2);
\node[block, fill=blue!15, right=0.75of start, anchor=west] (shaper) {Amplitude \\ Shaping \\ $a\in\calA$};
\path[line_arrow] (start) -- (shaper) node[pos=0.35,above] {$k$ bits};
\node[point, right=0.5of shaper] (amp2bit_helper){}; 
\node[point, right=3.5 of amp2bit_helper] (fec_helper){}; 
\path[line] (shaper) -- (amp2bit_helper) node[pos=0.95,above] {};
\node[block, below=1.25 of amp2bit_helper, anchor=center] (amp2bit) {Binary \\ Labeling};
\path[line_arrow] (amp2bit_helper) -- (amp2bit) node[pos=0.5,above] {};
\node[block, fill=red!20, below=1.6 of fec_helper, anchor=center, minimum height=15mm] (fec) {Systematic \\ FEC Coding \\ $\Rc = \frac{m+\gamma}{m+1}$};
\path[line_arrow] (amp2bit) -- (amp2bit-|fec.west) node[pos=0.5,above] {$\underline{c}_1\underline{c}_2\cdots\underline{c}_{m}$};
\path[line_arrow] (start_below) -- (start_below-|fec.west) node[pos=0.05,below] {$\gamma n$ bits};
\path[line_arrow,draw=none] (start_below) -- (start_below-|fec.west) node[pos=0.65,below] {$\us_{\text{i}} = (s_1, s_2,\cdots, s_{\gamma n})$};
\node[block_circ, draw=none, right=8.5 of start, anchor=center] (otimes) {};
\node[point, below=0.9 of otimes] (otimesbelow1){};
\node[point, below=1.7 of otimes] (otimesbelow2){};
\path[line_arrow] (amp2bit_helper) -- (otimes.center) node[pos=1.025,above,color=blue] {{$\ua = (a_1, a_2,\cdots, a_n)$}};
\path[line_arrow] (fec.east|-otimesbelow1) -- (otimesbelow1) node[pos=1.1,above,color=red] {{$\us_{\text{i}} = (s_1, s_2,\cdots, s_{\gamma n})$}};
\path[line_arrow] (fec.east|-otimesbelow2) -- (otimesbelow2) node[pos=1.25,above,color=red] {{$\us_{\text{p}} = (s_{\gamma n+1}s_{\gamma n+2}\cdots s_{n})$}};
\draw [decorate,decoration={brace,amplitude=5pt,mirror}]
 (10.5,-1.9) -- (10.5,-0.7) node[midway,xshift=3em,font=\footnotesize,color=red]{$\us = (\us_{\text{i}},\us_{\text{p}})$};
\end{tikzpicture}
}
\caption{Probabilistic amplitude shaping with transmission rate $R = k/n+\gamma$ bit/1D. 
}
\label{fig:shapingsigncoding_gen}
\end{figure}

PAS~\cite{BochererSS2015_ProbAmpShap} is a capacity-achieving CM strategy in which constellation shaping and FEC coding are combined as shown in Figure~\ref{fig:shapingsigncoding_gen}.
In PAS, first an amplitude shaping block maps $k$-bit information strings to $n$-amplitude shaped sequences $\ua = (a_1, a_2,\cdots, a_n)$ in an invertible manner.
These~amplitudes are drawn from a $2^{m}$-ary alphabet $\calA$.
The amplitude shaping block can be realized using constant composition distribution matching~\cite{SchulteB2016_CCDM}, multiset-partition distribution matching~\cite{Fehenberger2019_MPDM}, shell~mapping~\cite{Schulte2019_SMDM}, enumerative sphere shaping~\cite{GultekinHKW2019_ESSforShortWlessComm}, etc.

After $n$ amplitudes are generated, binary labels $\underline{c}_1\underline{c}_2\cdots \underline{c}_{m}$ of the amplitudes $\ua$ and an additional $\gamma n$-bit information string $\us_\text{i} = (s_1, s_2,\cdots, s_{\gamma n})$ are fed to a rate $(m+\gamma)/(m+1)$ systematic FEC encoder.
The encoder produces $(1-\gamma)n$ parity bits $\us_{\text{p}} = (s_{\gamma n+1}, s_{\gamma n+2},\cdots, s_n)$.
The additional data bits $\us_{\text{i}}$ and the parity bits $\us_{\text{p}}$ are used as the signs $\us = (s_1, s_2,\cdots, s_n)$ for the amplitudes $\ua$.
Finally, probabilistically shaped channel inputs $\ux = \us\otimes\ua$ are transmitted through the channel.
Here, $\gamma$ is the rate of the additional information in bits per symbol (bit/1D) or, equivalently, the~fraction of signs that are selected directly by data bits.
The transmission rate of PAS is $R = k/n + \gamma$ in bit/1D.

\subsection{Probabilistic Amplitude Shaping: Achievable~Rates}
Based on Gallager's error exponent approach~\cite[Ch. 5]{Gallager1968_ITbook}, AIRs of PAS were investigated in~\cite{Bocherer2018_AchvRates4ProbShap,Bocherer2018_PrinciplesCM,Amjad2018_InfRatesErrExponents}.
In~\cite{Bocherer2018_AchvRates4ProbShap}, a~random code ensemble was considered from which the channel inputs $\ux$ were drawn.
Then, the~AIR in~\cite[Eqs. (32)--(34)]{Bocherer2018_AchvRates4ProbShap} was derived for a general memoryless decoding metric $\mathbbm{q}(x,y)$.
It was shown that by properly selecting $\mathbbm{q}(x,y)$, $I(X;Y)$ and the rate \eqref{eq:rbmd} can be recovered from the derived AIR, and~consequently, they can be achieved with~PAS.

Computing error exponents for PAS was also the main concern of the work presented in~\cite[Ch. 10]{Bocherer2018_PrinciplesCM}.
The difference from~\cite{Bocherer2018_AchvRates4ProbShap} was in the random coding setup.
In~\cite[Ch. 10]{Bocherer2018_PrinciplesCM}, a~random code ensemble was considered from which only the {signs} $\us$ of the channel inputs were drawn at random.
We call this the {random sign-coding} setup.
The error exponent~\cite[Eq. (10.42)]{Bocherer2018_PrinciplesCM} was then derived again for a general memoryless decoding metric.
Error exponents of PAS have also been examined based on the joint source-channel coding (JSCC) setup in~\cite{Amjad2018_InfRatesErrExponents,Amjad2018_InfRatesErrExponents_arxiv_mdpi}.
Random sign-coding was considered in~\cite{Amjad2018_InfRatesErrExponents,Amjad2018_InfRatesErrExponents_arxiv_mdpi}, but~only with symbol-metric decoding (SMD) and only for the specific case where $\gamma=0$.

\subsection{Our~Contribution}\label{ssec:contribution}
In this work, we derive AIRs of PAS in a random sign-coding framework based on weak typicality~\cite[Secs. 3.1,~7.6 and~15.2]{CoverT2006_ElementsofInfoTheo}.
We first consider basic sign-coding in which amplitudes of the channel inputs are generated constructively while the signs are drawn from a randomly generated code.
Basic sign-coding corresponds to PAS with $\gamma=0$.
Then, we consider modified sign-coding in which only some of the signs are drawn from the random code while the remaining are chosen directly by information bits.
Modified sign-coding corresponds to PAS with $0<\gamma<1$.
We compute AIRs for both SMD and~BMD.

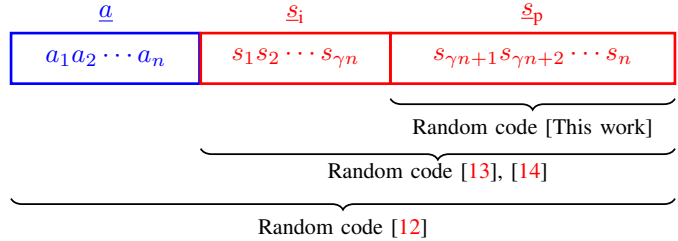
\begin{figure}[t]
\centering
\resizebox{0.5\columnwidth}{!}{
\begin{tikzpicture}[]
\draw[blue, very thick] (0,0) rectangle (2.98,0.8);
\node[color=blue] (amps) at (1.5,0.4) {\scalebox{1.2}{$a_1a_2\cdots a_n$}};
\node[color=blue] (amps) at (1.5,1.1) {\scalebox{1.2}{${\ua}$}};
\draw[red, very thick] (3,0) rectangle (6,0.8);
\node[color=red] (amps) at (4.5,0.4) {\scalebox{1.2}{$s_1s_2\cdots s_{\gamma n}$}};
\node[color=red] (amps) at (4.5,1.1) {\scalebox{1.2}{${\us_{\text{i}}}$}};
\draw[red, very thick] (6,0) rectangle (10.5,0.8);
\node[color=red] (amps) at (8.25,0.4) {\scalebox{1.2}{$s_{\gamma n +1}s_{\gamma n+2}\cdots s_{ n}$}};
\node[color=red] (amps) at (8.25,1.1) {\scalebox{1.2}{${\us_{\text{p}}}$}};
\draw [decorate,thick,decoration={brace,amplitude=5pt,mirror,raise=1.5ex}]
 (6,0) -- (10.5,0) node[midway,yshift=-2em]{Random code [This work]};
\draw [decorate,thick,decoration={brace,amplitude=5pt,mirror,raise=6.5ex}]
 (3,0) -- (10.5,0) node[midway,yshift=-4em]{Random code~\cite{Bocherer2018_PrinciplesCM,Amjad2018_InfRatesErrExponents}};
\draw [decorate,thick,decoration={brace,amplitude=5pt,mirror,raise=11.5ex}]
 (0,0) -- (10.5,0) node[midway,yshift=-6.5em]{Random code~\cite{Bocherer2018_AchvRates4ProbShap}};
\end{tikzpicture}
}
\caption{The scope of the random coding experiments considered in this work and in~\cite{Bocherer2018_AchvRates4ProbShap,Bocherer2018_PrinciplesCM,Amjad2018_InfRatesErrExponents}.}
\label{fig:randomcodeextents}
\end{figure}

Our first objective is to provide alternative proofs of achievability in which the codes are generated as constructively as possible.
In our random sign-coding experiment, both the amplitude sequences ($\ua$) and the sign sequence parts ($\us_{\text{i}}$) that are information bits are constructively produced, and~only the remaining signs ($\us_{\text{p}}$) are randomly generated as illustrated in Figure~\ref{fig:randomcodeextents}.
In most proofs of Shannon's channel coding theorem, channel input sequences ($\ux$) are drawn at random, and~the existence of a good code is demonstrated. Therefore, these proofs are not constructive and cannot be used to identify good codes as discussed, e.g.,~in~\cite[Sec. I]{Shulman1999_RandomCodingforRandomCodes} and the references therein.
On the other hand, in~our proofs using random sign-coding arguments, it is self-evident how---at least a part of---the code should be constructed.
Our second objective is to provide a unified framework in which all possible PAS scenarios are considered, i.e.,~SMD or BMD at the receiver with $0\leq \gamma <1$, and~corresponding AIRs are determined using a single technique, i.e.,~the random sign-coding~argument.

Note that our approach differs from the random sign-coding setup considered in~\cite{Bocherer2018_PrinciplesCM,Amjad2018_InfRatesErrExponents} where {all} signs ($\us_{\text{i}}$ and $\us_{\text{p}}$) were generated randomly, which was called partially systematic encoding in~\cite[Ch. 10]{Bocherer2018_PrinciplesCM}. 
We will show later that only $\us_{\text{p}}$ needs to be chosen randomly.
Furthermore, we define a special type of typicality ($\calB$-typicality; see Definition~\ref{def:Beps} below) that allows us to avoid the mismatched JSCC approach of~\cite{Amjad2018_InfRatesErrExponents}.

\section{Preliminaries}\label{sec:prelim}
\vspace{-6pt}
\subsection{Memoryless~Channels}
We consider communication over a memoryless channel with discrete input $X\in\calX$ and discrete output $Y\in\calY$.
The channel law is given by:
\begin{equation}
 p(\uy|\ux) = \prod_{i=1}^n p(y_i|x_i).
\end{equation}
Later in Example~\ref{ex:awgnsigncoding}, we will also discuss the AWGN channel $Y = X + Z$ where $Z$ is zero-mean Gaussian with variance $\sigma^2$.
In this case, we assume that the channel output $Y$ is a quantized version of the continuous channel output $X+Z$.
Furthermore, we assume that this quantization has a resolution high enough that the discrete-output channel is an accurate model for the underlying continuous-output channel.
Therefore, the~achievability results we will obtain for discrete memoryless channels carry over to the discrete-input AWGN~channel.

\subsection{Typical~Sequences}
We will provide achievability proofs based on weak typicality.
In this section, which is based on~\cite[Secs.~3.1,~7.6, and~15.2]{CoverT2006_ElementsofInfoTheo}, we formally define weak typicality and list its properties that will be used in this~paper.

Let $\varepsilon>0$ and $n$ be a positive integer.
Consider the random variable $X$ with probability distribution $p(x)$.
Then, the~(weak) typical set $\Aeps(X)$ of length-$n$ sequences with respect to $p(x)$ is defined as:
\begin{equation}
\Aeps(X) \triangleq \left\{ \ux\in\calX^n : \left|-\frac{1}{n} \log p(\ux) -H(X)\right| \leq \varepsilon \right\}, \label{eq:typicalset}
\end{equation}
where:
\begin{equation}
p(\ux) \triangleq \prod_{i=1}^n p(x_i).
\end{equation}
The cardinality of the typical set $\Aeps(X)$ satisfies~\cite[Thm. 3.1.2]{CoverT2006_ElementsofInfoTheo}:
\begin{equation}
(1-\varepsilon)2^{n(H(X)-\varepsilon)} \stackrel{\text{(a)}}{\leq} \left|\Aeps(X)\right| \stackrel{\text{(b)}}{\leq} 2^{n(H(X)+\varepsilon)}, \label{eq:typcardi}
\end{equation}
where (a) holds for $n$ sufficiently large and (b) holds for all $n$.
For $\ux\in\Aeps(X)$, the~probability of occurrence can be bounded as~\cite[Eq. (3.6)]{CoverT2006_ElementsofInfoTheo}:
\begin{equation}
2^{-n(H(X)+\varepsilon)} \leq p(\ux) \leq 2^{-n(H(X)-\varepsilon)}. \label{eq:typprob}
\end{equation}

The idea of typical sets can be generalized for pairs of $n$-sequences.
Now, consider the pair of random variables $(X,Y)$ with probability distribution $p(x,y)$.
Then, the~typical set $\Aeps(XY)$ of pairs of length-$n$ sequences with respect to $p(x,y)$ is defined as:
\begin{IEEEeqnarray}{rCl} 
\Aeps(XY) \triangleq \bigg\{ (\ux,\uy)\in\calX^n\times\calY^n : && \left|-\frac{1}{n} \log p(\ux) -H(X)\right| \leq \varepsilon, \nonumber \\ && \left|-\frac{1}{n} \log p(\uy) -H(Y)\right| \leq \varepsilon, \nonumber \\ && \left|-\frac{1}{n} \log p(\ux,\uy) -H(X,Y)\right| \leq \varepsilon \bigg\} \label{eq:jointlytypicalset} 
\end{IEEEeqnarray}
where:
\begin{equation}
p(\ux,\uy) \triangleq \prod_{i=1}^n p(x_i,y_i),
\end{equation}
and where $p(x)$ and $p(y)$ are the marginal distributions that correspond to $p(x,y)$.
The cardinality of the typical set $\Aeps(XY)$ satisfies~\cite[Thm. 7.6.1]{CoverT2006_ElementsofInfoTheo}:
\begin{equation}
\left|\Aeps(XY)\right| \leq 2^{n(H(X,Y)+\varepsilon)} \label{eq:jointtypcardi}
\end{equation}
for all $n$.
For $(\ux,\uy)\in\Aeps(XY)$, the~probability of occurrence can be bounded in a similar manner to \eqref{eq:typprob} as:
\begin{equation}
2^{-n(H(X,Y)+\varepsilon)} \leq p(\ux,\uy) \leq 2^{-n(H(X,Y)-\varepsilon)}. \label{eq:jointtypprob}
\end{equation}
Along the same lines, joint typicality can be extended for collections of $n$-sequences $(\uX_1, \uX_2,\cdots, \uX_m)$ and the corresponding typical set $\Aeps(X_1X_2\cdots X_m)$ can be defined similar to how \eqref{eq:typicalset} was extended to \eqref{eq:jointlytypicalset}.
Then, for~$(\ux_1, \ux_2,\cdots, \ux_m)\in\Aeps(X_1X_2\cdots X_m)$, the~probability of occurrence can be bounded in a similar manner to \eqref{eq:jointtypprob} as:
\begin{equation}
2^{-n(H(\boldX)+\varepsilon)} \leq p(\ux_1, \ux_2,\cdots, \ux_m) \leq 2^{-n(H(\boldX)-\varepsilon)},\label{eq:3jointtypprob}
\end{equation}
where $\boldX = (X_1, X_2, \dots, X_m)$.

Finally, we fix $\ux$.
The conditional (weak) typical set $\Aeps(Y|\ux)$ of length-$n$ sequences is defined as:
\begin{equation}
\Aeps(Y|\ux) = \left\{\uy: (\ux,\uy) \in \Aeps(XY)\right\}.
\end{equation}
In other words, $\Aeps(Y|\ux)$ is the set of all $\uy$ sequences that are jointly typical with $\ux$.
For $\ux \in \Aeps(X)$ and for sufficiently large $n$, the~cardinality of the conditional typical set $\Aeps(Y|\ux)$ satisfies~\cite[Thm. 15.2.2]{CoverT2006_ElementsofInfoTheo}:
\begin{equation}
|\Aeps(Y|\ux)| \leq 2^{n(H(Y|X)+2\varepsilon)}. \label{eq:condtypcardi}
\end{equation}

\begin{definition}[{$\calB$-typicality}]\label{def:Beps}
Let the input probability distribution $p(u)$ together with the transition probability distribution $p(v|u)$ determine the joint probability distribution $p(u,v)= p(u)p(v|u)$. 
Now, we define:
\begin{equation} \label{eq:Bset}
\BepsV(U) \define \Big\{ \uu : \uu \in \Aeps(U) \text{ and } \Pr\big\{ (\uu,\uV) \in \Aeps(UV) \mid \uU=\uu) \big\} \geq 1-\varepsilon \Big\}, 
\end{equation}
where $\uV$ is the output sequence of a ``channel'' $p(v|u)$ when sequence $\uu$ is input. 
\end{definition}

The set $\BepsV(U)$ in \eqref{eq:Bset} guarantees that a sequence $\uu$ in this $\calB$-typical set will with high probability lead to a sequence $\uv$ that is jointly typical with $\uu$. 
We note that $U$ and/or $V$ can be composite.
The set $\BepsV(U)$ has three properties, as~stated in Lemma~\ref{Lemma.Ps}, the~proof of which is given in Appendix~\ref{App.P2}.

\begin{lemma}[{$\calB$-typicality properties}]\label{Lemma.Ps}
The set $\BepsV(U)$ in Definition~\ref{def:Beps} has the following~properties:
\begin{enumerate}[leftmargin=5ex,itemsep=0.5ex,label={$P_{\arabic*}$}]
\item : \label{prop:weaktyp1}
For $\uu \in \BepsV(U)$,
\begin{equation}
 2^{-n(H(U)+\varepsilon) } \leq p(\uu) \leq 2^{-n(H(U)-\varepsilon)}.
\end{equation}

\item : \label{prop:weaktyp2}
For $n$ large enough,
\begin{equation}
 \sum_{\uu \notin \BepsV(U)} p(\uu) \leq \varepsilon. \nonumber
\end{equation}

\item : \label{prop:weaktyp3} 
$|\BepsV(U)| \leq 2^{n(H(U)+\varepsilon)}$ holds for all $n$, while $|\BepsV(U)| \geq (1-\varepsilon)2^{n(H(U)-\varepsilon)}$ holds for $n$ large enough.
\end{enumerate}
\end{lemma}

\section{Random Sign-Coding~Experiment}\label{sec:randomsigncoding}
We consider $2^{m+1}$-ary amplitude shift keying ($M$-ASK) alphabets \mbox{$\calX=\{-M+1, -M+3, \cdots, M-1\}$} where $M=2^{m+1}$.
We note that $\calX$ is symmetric around the origin and can be factorized as $\calX = \calS \calA$.
Here, $\calS=\{-1,+1\}$ and $\calA = \{+1,+3,\cdots,M-1\}$ are the sign and amplitude alphabets, respectively.
Accordingly, any channel input $x\in\calX$ can be written as the multiplication of a sign and an amplitude, i.e.,~$x = s \otimes a$.

\subsection{Random Sign-Coding~Setup} 
We cast the PAS structure shown in Figure~\ref{fig:shapingsigncoding_gen} as a {sign-coding} structure as in Figure~\ref{fig:shapingsigncoding}.
The~sign-coding setup consists of two layers: a shaping layer and a coding~layer.

\begin{definition}[{Sign-coding}]\label{def:signcoding}

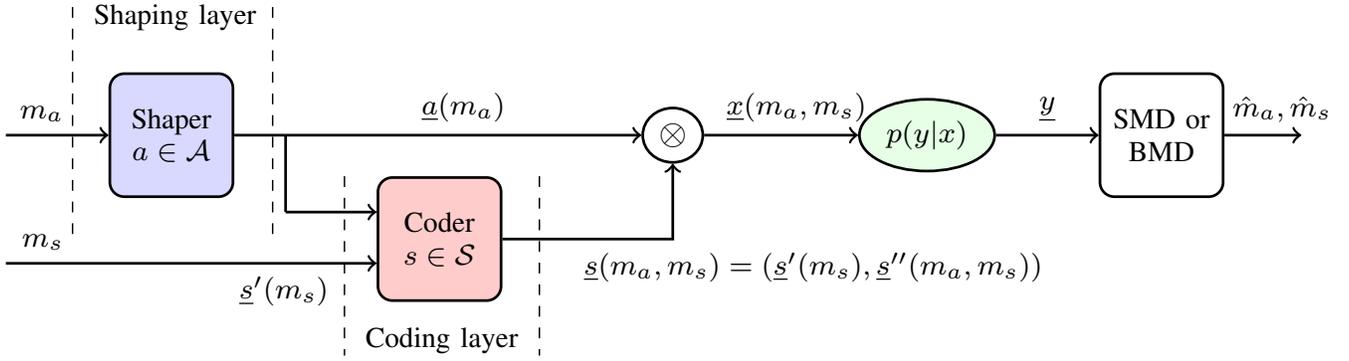
\begin{figure}[t]
\centering
\resizebox{1\columnwidth}{!}{
\begin{tikzpicture}[
line width=0.75pt,
block/.style={rectangle, rounded corners,draw,inner sep=2pt,minimum width=5mm, minimum height=5mm,font=\footnotesize,align=center},
block_circ/.style={ellipse, rounded corners,draw,inner sep=2pt,minimum width=5mm, minimum height=5mm,align=center},
block_ell/.style={ellipse, rounded corners,draw,inner sep=2pt,minimum width=10mm, minimum height=7mm,font=\footnotesize,align=center},
line_arrow/.style = {draw,->,font=\footnotesize},
line/.style = {draw,-,font=\footnotesize}
]
\pgfmathsetmacro{\nodedist}{1}
\coordinate (start) at (0,0);
\coordinate (start_below) at (0,-1.25);
\node[block, fill=blue!15, right=of start, minimum height=12mm, minimum width=12mm, anchor=west] (shaper) {Shaper \\ $a\in\calA$};
\path[line_arrow] (start) -- (shaper) node[pos=0.35,above] {$m_a$};
\node[point, right=0.5 of shaper] (amp2bit_helper){}; 
\node[point, right=1.5 of amp2bit_helper] (fec_helper){}; 
\node[point, below=0.75 of amp2bit_helper] (amp2bit_helper_below){}; 
\path[line] (shaper) -- (amp2bit_helper) node[pos=0.95,above] {};
\node[block, fill=red!20, below=0.4 of fec_helper, anchor=north, minimum height=12mm, minimum width=12mm] (fec) {Coder \\ $s\in\calS$};
\path[line] (amp2bit_helper) -- (amp2bit_helper_below) node[pos=0.5,above] {};
\path[line_arrow] (amp2bit_helper_below) -- (amp2bit_helper_below-|fec.west) node[pos=0.5,above] {};
\path[line_arrow] (start_below) -- (start_below-|fec.west) node[pos=0.1,above] {$m_s$};
\path[line_arrow,draw=none] (start_below) -- (start_below-|fec.west) node[pos=0.75,below] {$\use(m_s)$};
\node[block_circ, right=6.5 of start, anchor=center] (otimes) {\scalebox{1}{$\otimes$}};
\path[line_arrow] (fec.east) -| (otimes.south) node[pos=0.2, below right] {$\us(m_a,m_s) = (\use(m_s), \ust(m_a,m_s))$};
\path[line_arrow,draw=none] (fec.east) -| (otimes.south) node[pos=0.85,left] {};
\path[line_arrow] (amp2bit_helper) -- (otimes) node[pos=0.5,above]
{$\ua(m_a)$};
\node[block_ell, fill=green!10, right=1.5 of otimes.east, anchor=west] (chan) {$p(y|x)$};
\path[line_arrow] (otimes) -- (chan) node[pos=0.6,above] {$\ux(m_a,m_s)$};
\node[block, right=1 of chan.east, anchor=west, minimum width = 12mm, minimum height = 12mm] (decode) {SMD or \\ BMD};
\path[line_arrow] (chan) -- (decode) node[pos=0.5,above] {$\uy$};
\node[point, right=0.75 of decode] (end){};
\path[line_arrow] (decode) -- (end) node[pos=0.75,above] {$\hat{m}_a,\hat{m}_s$};
\path[line,dashed,thin] (0.65,1.25) -- (0.65,-1) node[pos=0.5,above] {};
\path[line,dashed,thin] (2.6,1.25) -- (2.6,-1) node[pos=0.5,above] {};
\path[line,dashed,thin] (3.3,-0.4) -- (3.3,-2.15) node[pos=0.5,above] {};
\path[line,dashed,thin] (5.2,-0.4) -- (5.2,-2.15) node[pos=0.5,above] {};
\node[block, draw=none] (decode) at (1.65,1.15) {Shaping layer};
\node[block, draw=none] (decode) at (4.25,-2) {Coding layer};
\end{tikzpicture}
}
\caption{Sign-coding structure: sign-coding (coder) is combined with amplitude shaping (shaper). SMD, symbol-metric decoding; BMD, bit-metric~decoding.}
\label{fig:shapingsigncoding}
\end{figure}

For every message index pair $(m_a,m_s)$, with~uniform $m_a\in\{1,2,\cdots,M_a\}$ and uniform $m_s\in\{1,2,\cdots,M_s\}$, a~sign-coding structure as shown in Figure~\ref{fig:shapingsigncoding} consists of the~following.
\begin{itemize}
\item A {shaping layer} that produces for every message index $m_a$, a~length-$n$ shaped amplitude sequence $\ua(m_a)$ where the mapping is one-to-one. 
The set of amplitude sequences is assumed to be shaped, but~uncoded.
\item An additional $n_1$-bit (uniform) information string in the form of a sign sequence part \mbox{$\use(m_s) = (s_{1}(m_s), s_{2}(m_s),\cdots, s_{n_1}(m_s))$} for every message index $m_s$.
\item A {coding layer} that extends the sign sequence part $\use(m_s)$ by adding a second (uniform) sign sequence part $\ust(m_a,m_s) = (s_{n_1+1}(m_a,m_s),s_{n_1+2}(m_a,m_s),\cdots,s_{n}(m_a,m_s))$ of length-$n_2$ for all $m_a$ and $m_s$. 
This is obtained by using an encoder that produces redundant signs in the set $\calS$ from $\ua(m_a)$ and $\use(m_s)$. 
Here, $n_1+n_2=n$. 
\end{itemize}
Finally, the~transmitted sequence is $\ux(m_a,m_s) = \ua(m_a) \otimes \us(m_a,m_s)$, where~\mbox{$\us(m_a,m_s)= (\use(m_s), \ust(m_a,m_s))$}.
The sign-coding setup with $n_1=0$ ($\gamma=0$) is called {basic sign-coding}, while the setup with $n_1>0$ ($\gamma>0$) is called {modified sign-coding}.
\end{definition}

\subsection{Shaping~Layer} 
\begin{figure}[t]
\centering
\resizebox{0.43\columnwidth}{!}{
\begin{tikzpicture}[
line width=0.75pt,
block/.style={rectangle, rounded corners,draw,inner sep=2pt,minimum width=5mm, minimum height=5mm,font=\footnotesize,align=center},
block_circ/.style={ellipse, rounded corners,draw,inner sep=2pt,minimum width=5mm, minimum height=5mm,align=center},
block_ell/.style={ellipse, rounded corners,draw,inner sep=2pt,minimum width=10mm, minimum height=7mm,font=\footnotesize,align=center},
line_arrow/.style = {draw,->,font=\footnotesize},
line/.style = {draw,-,font=\footnotesize}
]
\coordinate (start) at (0,0);
\node[block,fill=blue!15, right= of start, minimum height=12mm, minimum width=12mm] (shaplayer) {Shaper \\ $p(a), a\in\calA$};
\path[line_arrow] (start) -- (shaplayer) node[pos=0.5,above] {$m_a$};
\node[point, right=0.75 of shaplayer.east] (shaplayer_r){}; 
\path[line_arrow] (shaplayer) -- (shaplayer_r) node[pos=0.5,above right] {$\ua(m_a)\in\BepsSY(A)$};
\end{tikzpicture}
}
\caption{Shaping layer of the random sign-coding setup with~SMD.}
\label{fig:smdshapinglayer}
\end{figure}
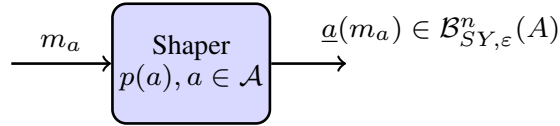

When SMD is employed at the receiver, the~shaping layer is as shown in Figure~\ref{fig:smdshapinglayer}.
Here, let $A$ be distributed with $p(a)$ over $a\in\calA$.
Then, the~shaper produces for every message index $m_a$ a length-$n$ amplitude sequence $\ua(m_a)\in\BepsSY(A)$.
{We note that for this sign-coding setup, the~rate is:
\begin{IEEEeqnarray}{rCl} \label{eq:signcodingrateexpression}
R &=& \frac{1}{n} \log_2 |M_aM_s| = \gamma + \frac{1}{n} \log_2 |\BepsSY(A)| \geq H(A) + \gamma -2\varepsilon 
\end{IEEEeqnarray}
where the inequality in \eqref{eq:signcodingrateexpression} follows for $n$ large enough from~\ref{prop:weaktyp3}.}

\begin{figure}[t]
\centering
\resizebox{0.6\columnwidth}{!}{
\begin{tikzpicture}[
line width=0.75pt,
block/.style={rectangle, rounded corners,draw,inner sep=2pt,minimum width=5mm, minimum height=5mm,font=\footnotesize,align=center},
block_circ/.style={ellipse, rounded corners,draw,inner sep=2pt,minimum width=5mm, minimum height=5mm,align=center},
block_ell/.style={ellipse, rounded corners,draw,inner sep=2pt,minimum width=10mm, minimum height=7mm,font=\footnotesize,align=center},
line_arrow/.style = {draw,->,font=\footnotesize},
line/.style = {draw,-,font=\footnotesize}
]
\pgfmathsetmacro{\nodedist}{1}
\coordinate (start) at (0,0);
\node[block, fill=blue!15,right=0.75 of start, minimum height=12mm, minimum width=12mm, anchor=west] (shaplayer) {Shaper \\ $p(\boldb)$, \\ $\boldb\in\{0,1\}^m$};
\path[line_arrow] (start) -- (shaplayer) node[pos=0.5,above] {$m_a$};
\node[block,fill=blue!15, right=4.2 of shaplayer, minimum height=12mm, minimum width=12mm, anchor=west] (mapper) {Symbol-wise \\ Mapping \\ $f(\boldb)$};
\path[line_arrow] (shaplayer) -- (mapper) node[pos=0.5,above] {$\uboldb(m_a) \in \BepsSY(B_1 B_2 \cdots B_m)$};
\node[point, right=0.75 of mapper] (end){};
\path[line_arrow] (mapper) -- (end) node[pos=0.6,above] {$\ua(m_a)$};
\end{tikzpicture}
}
\caption{Shaping layer of the random sign-coding setup with BMD for~M-ASK.}
\label{fig:bmdshapinglayer}
\end{figure}
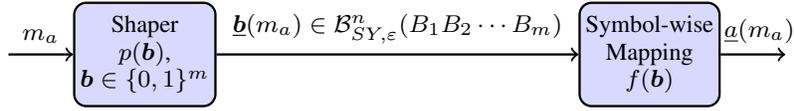

On the other hand, when BMD is used at the receiver, the~shaping layer is as shown in Figure~\ref{fig:bmdshapinglayer}.
Here, let $ \boldB = (B_1,B_2,\cdots, B_{m})$ be distributed with $p(\boldb) = p(b_1,b_2,\cdots,b_{m})$ over $(b_1,b_2,\cdots,b_{m})\in\{0,1\}^{m}$.
The shaper produces for every message index $m_a$ an $n$-sequence of $m$-tuples $\uboldb(m_a) = (\ub_1(m_a),\ub_2(m_a),\cdots,\ub_{m}(m_a))\in\BepsSY(B_1B_2\cdots B_{m})$.
Then, each $m$-tuple is mapped to an amplitude sequence $\ua(m_a)$ by a symbol-wise mapping function $f(\cdot)$.
{We note that for this sign-coding setup, the~rate is:
\begin{IEEEeqnarray}{rCl} \label{eq:signcodingrateexpression2}
R &=& \frac{1}{n} \log_2 |M_aM_s| = \gamma + \frac{1}{n} \log_2 |\BepsSY(\boldB)| \geq H(\boldB) + \gamma -2\varepsilon 
\end{IEEEeqnarray}
where the inequality in \eqref{eq:signcodingrateexpression2} follows for $n$ large enough from~\ref{prop:weaktyp3}.}

To realize $f(\cdot)$, we label the channel inputs with $(m+1)$-bit strings.
The amplitude is addressed by $m$ amplitude bits $(B_1,B_2,\cdots, B_{m})$, while the sign is addressed by a sign bit $S$.
The symbol-wise mapping function $f(\cdot)$ in Figure~\ref{fig:bmdshapinglayer} uses the addressing $(B_1,B_2,\cdots, B_{m}) \Longleftrightarrow A$.
We emphasize that unlike the case in Section~\ref{sec:AIRs}, we use $(S,B_1,B_2,\cdots, B_{m})$ to denote a channel input instead of $(C_1,C_2,\cdots, C_{m+1})$.
Amplitudes and signs of $x\in\calX$ are tabulated for 8-ASK in Table~\ref{tab:8ask} along with an example of the mapping function $f(b_1,b_2)$, namely the binary reflected Gray code~\cite[Defn. 2.10]{Szczecinski2015_BICMbook}.

\begin{table}[h]
\caption{Input alphabet and mapping function for 8-ASK.}
\begin{center}
\renewcommand{\arraystretch}{1.2}
\scalebox{1.2}{
\begin{tabular}{c|cccccccc} 
\textcolor{red}{$A$} & 7 & 5 & 3 & 1 & 1 & 3 & 5 & 7 \\
\textcolor{blue}{$S$} & -1 & -1 & -1 & -1 & 1 & 1 & 1 & 1 \\
\hline\hline
$X$ & -7 & -5 & -3 & -1 & 1 & 3 & 5 & 7 \\
\hline\hline
\textcolor{red}{$B_1$} & 0 & 0 & 1 & 1 & 1 & 1 & 0 & 0 \\ 
\textcolor{red}{$B_2$} & 0 & 1 & 1 & 0 & 0 & 1 & 1 & 0 \\ 
\end{tabular}
}
\label{tab:8ask}
\end{center}
\end{table}

\subsection{Decoding~Rules}\label{sec:decoding}
At the receiver, SMD finds the unique message index pair $(\hat{m}_a,\hat{m}_s)$ such that the corresponding amplitude-sign sequence is jointly typical with the received output sequence $\uy$, i.e.,~$(\ua(\hat{m}_a),\us(\hat{m}_a,\hat{m}_s),\uy) \in \Aeps(ASY)$.

On the other hand, BMD finds the unique message index pair $(\hat{m}_a,\hat{m}_s)$ such that the corresponding bit and sign sequences are (individually) jointly typical with the received output sequence $\uy$, i.e.,~$(\us(\hat{m}_a,\hat{m}_s),\uy) \in \Aeps(SY)$ and $(\ub_j(\hat{m}_a),\uy) \in \Aeps(B_j Y)$ for $j = 1, 2,\cdots, m$.
We note that the decoder can use bit metrics $p(b_{ji}=1|y_i)=1-p(b_{ji}=0|y_i)$ for $j = 1, 2,\cdots, m$ and $i=1,2,\cdots,n$ to find $p(\ub_j|\uy)$.
Here, $b_{ji}$ is the $j^{\text{th}}$ bit of the $i^{\text{th}}$ symbol.
Together with $p(\uy)$ and $p(\ub_j)$, the~decoder can check whether $(\ub_j,\uy) \in \Aeps(B_j Y)$. 
We note that $B_j$ is in general not uniform. 
A similar statement holds for the uniform sign $S$.

\section{Achievable Information Rates of~Sign-Coding}\label{sec:airsofsigncoding}
Here, we investigate AIRs of the sign-coding architecture in Figure~\ref{fig:shapingsigncoding}.
We consider both SMD and BMD at the receiver. 
In what follows, four AIRs are presented. 
The proofs are based on $\calB$-typicality, a~variation of weak typicality, and~random sign-coding arguments and are given in Appendix~\ref{sec:app_thms}.
As~indicated in Definition~\ref{def:signcoding}, signs $S$ are assumed to be uniform in the proofs.
We have not applied weak typicality for continuous random variables, discussed in~\cite[Sec. 8.2]{CoverT2006_ElementsofInfoTheo} and~\cite[Sec. 10.4]{Yeung2008_ITandNworkCode}, since~our channels are discrete-input. 
However, it is also possible to develop a hybrid version of weak typicality that matches with discrete-input continuous-output~channels.

In the following, the~concept of AIR is formally defined in the sign-coding context.
\begin{definition}[{Achievable information rate}]
A rate $R$ is said to be achievable if for every $\delta>0$ and $n$ large enough, there exists a sign-coding encoder and a decoder such that $(1/n)\log_2\left({M_aM_s}\right) \geq R-\delta$ and error probability $\Pe \leq \delta$.
\end{definition} 

\subsection{Sign-Coding with Symbol-Metric~Decoding}
\begin{theorem}[{Basic sign-coding with SMD}]\label{bsc_smd}
For a memoryless channel $\{ \calX, p(y|x), \calY \}$ with amplitude shaping and basic sign-coding, the~rate:
\begin{equation}
R_{\text{SMD}}^{\gamma=0} = \max_{p(a): H(A)\leq I(SA;Y)} H(A) \label{R.bas}
\end{equation}
is achievable using SMD.
\end{theorem}

\begin{figure}[t]
\centering
\resizebox{0.7\columnwidth}{!}{\includegraphics{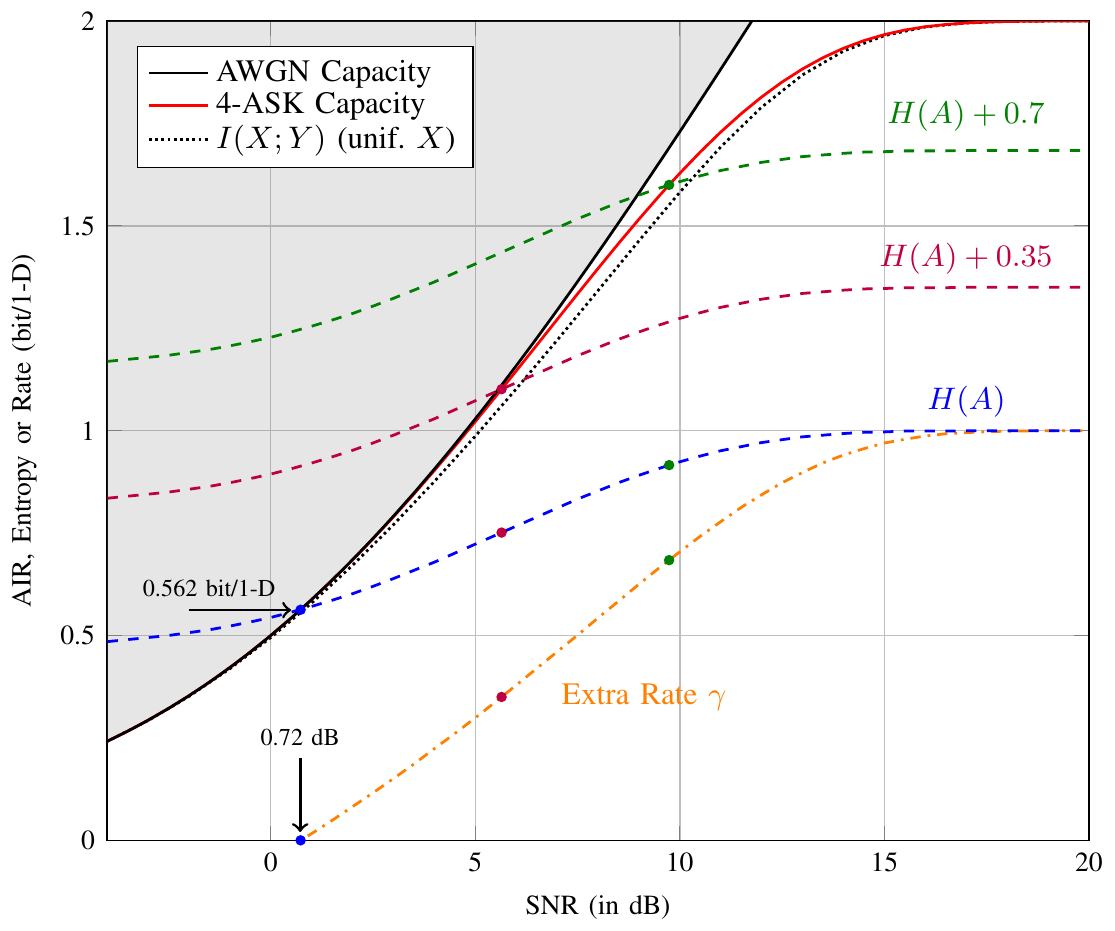}}
\caption{Sign-coding with SMD for 4-ASK. All $C_{\text{4-ASK}}\geq 0.562$~bit/1D can be achieved with sign-coding. AIR, achievable information~rate.}
\label{fig:4ask}
\end{figure}

Theorem~\ref{bsc_smd} implies that for a memoryless channel, the~rate $R=H(A)$ is achievable with basic sign-coding, as~long as $H(A) \leq I(SA;Y) = I(X;Y)$ is satisfied.
For the AWGN channel, this means that a range of rate-SNR pairs are achievable.
Here, SNR denotes the signal-to-noise ratio.
One of these points, $H(A)=I(SA;Y)$, is on the capacity-SNR curve.
Note that here, ``capacity'' indicates the largest achievable rate using $\calX$ as the channel input alphabet under the average power constraint.
{It can be observed from Figure~\ref{fig:4ask} discussed in Example~\ref{ex:awgnsigncoding} that there indeed exists an amplitude distribution $p(a)$ for which $H(A)=I(SA;Y)$.}

\begin{theorem}[{Modified sign-coding with SMD}]\label{mod_smd}
For a memoryless channel $\{\calX, p(y|x), \calY\}$ with amplitude shaping and modified sign-coding, the~rate:
\begin{equation}
R_{\text{SMD}}^{\gamma>0} = \max_{p(a), \gamma: H(A)+\gamma\leq I(SA;Y)} H(A)+\gamma \label{R.smd.gen}
\end{equation}
is achievable using SMD for $\gamma < 1$. 
\end{theorem}

Theorem~\ref{mod_smd} implies that for a memoryless channel, the~rate $H(A)+\gamma$ is achievable with modified sign-coding, as~long as $R=H(A)+\gamma \leq I(SA;Y)=I(X;Y)$ is satisfied.
For the AWGN channel, this means that all points on the capacity-SNR curve for which $H(X|Y) \leq 1-\gamma$ are achievable.
\mbox{This follows from}:
\begin{equation}
H(A)+\gamma \leq I(SA;Y) = H(SA) - H(SA|Y) = H(A) + 1 - H(X|Y), \label{eq:onebit}
\end{equation}
i.e., the~constraint in the maximization in \eqref{R.smd.gen}.

\begin{example}\label{ex:awgnsigncoding}
We consider the AWGN channel with average power constraint $E[X^2]\leq P$.
Figure~\ref{fig:4ask} shows the capacity of 4-ASK:
\begin{equation}
 C_{\text{4-ASK}} = \max_{\substack{p(x): \calX = \{-3, -1, +1, +3\}, \\ E\left[X^2\right]\leq P}} I(X;Y) \label{eq:4askcap}
\end{equation}
together with the amplitude entropy $H(A)$ of the distribution that achieves this capacity. 
Here, \mbox{$\snr = E[X^2]/\sigma^2$}, and~$\sigma^2$ is the noise variance.
Basic sign-coding achieves capacity only for $\SNR=0.72$~dB, i.e.,~at the point where $H(A)=I(X;Y)$, which is $C_{\text{4-ASK}}=0.562$~bit/1D.
We see from Figure~\ref{fig:4ask} that the shaping gap is negligible around this point, i.e.,~the capacity $C_{\text{4-ASK}}$ of 4-ASK and the MI $I(X;Y)$ for uniform $p(x)$ are virtually the same.
On the other hand, this gap is significant for larger rates, e.g.,~it is around 0.42~dB at 1.6 bit/1D.
To achieve rates larger than 0.562 bit/1D on the capacity-SNR curve, modified sign-coding ($\gamma>0$) is required.
At a given SNR, $C_{\text{4-ASK}}$ can be written as $C_{\text{4-ASK}} = H(A)+\gamma$, i.e.,~when the $H(A)$ curve is shifted above by $\gamma$, the~crossing point is again at $C_{\text{4-ASK}}$ for that SNR.
We also plot the additional rate $\gamma = C_{\text{4-ASK}}-H(A)$ in Figure~\ref{fig:4ask}. 
As an example, at~$\text{SNR}=9.74$~dB, $C_{\text{ASK}}=H(A)+\gamma=1.6$ can be achieved with modified sign-coding where $H(A)=0.9$ and $\gamma=0.7$.
We observe that sign-coding achieves the capacity of 4-ASK for $\SNR \geq 0.72$~dB.
\end{example}

\unskip

\subsection{Sign-Coding with Bit-Metric~Decoding}
The following theorems give AIRs for sign-coding with~BMD.

\begin{theorem}[{Basic sign-coding with BMD}]\label{bsc_bmd}
For a memoryless channel $\{\calX, p(y|x), \calY\}$ with amplitude shaping using $M$-ASK and basic sign-coding, the~rate:
\begin{equation}
R_{\text{BMD}}^{\gamma=0} = \max_{ \substack{p(\boldb): H(\boldB)\leq \rbmd(p(x))}} H(\boldB) \label{rbmd_bas}
\end{equation}
is achievable using BMD.
Here, $\boldB = (B_1, B_2,\dotsc, B_m)$, $p(\boldb) = p(b_1, b_2,\dotsc, b_m)$, and~\mbox{$p(x) = p(s, b_1, b_2, \dotsc, b_m)$}, and~$\rbmd(p(x))$ is as defined in \eqref{eq:rbmd}.
\end{theorem}

\begin{theorem}[{Modified sign-coding with BMD}]\label{mod_bmd}
For a memoryless channel $\{\calX, p(y|x), \calY\}$ with amplitude shaping using $M$-ASK and modified sign-coding, the~rate:
\begin{equation}
R_{\text{BMD}}^{\gamma>0} = \max_{ \substack{p(\boldb), \gamma: H(\boldB)+\gamma\leq \rbmd(p(x))}} H(\boldB) +\gamma \label{rbmd_bas}
\end{equation}
is achievable using BMD for $\gamma < 1$.
\end{theorem}

Theorems~\ref{bsc_bmd} and~\ref{mod_bmd} imply that for a memoryless channel, the~rate $R = H(\boldB)+\gamma = H(A)+\gamma$ is achievable with sign-coding and BMD, as~long as $R\leq\rbmd$ is satisfied.

\begin{remark}[{Random sign-coding with binary linear codes}]
An amplitude can be represented by $m$ bits. 
We can uniformly generate a code matrix with $mn$ rows of length $n$. 
This matrix can be used to produce the sign sequences.
This results in the pairwise independence of any two different sign sequences, as~is explained in the proof of~\cite[Thm.~6.2.1]{Gallager1968_ITbook}. 
Inspection of the proof of our Theorem~\ref{bsc_smd} shows that only the pairwise independence of sign sequences is needed. 
Therefore, achievability can also be obtained with a binary linear code.
Note that our linear code can also be seen as a systematic code that generates parity. 
The code rate of the corresponding systematic code is $m/(m+1)$.
For BMD, a~similar reasoning shows that linear codes lead to achievability, and~also for modified sign-coding, achievability follows for binary linear codes.
The rate of the systematic code that corresponds to the modified setting is $(m+\gamma)/(m+1)$.
\end{remark}

\section{Conclusions}\label{sec:conc}
In this paper, we studied achievable information rates (AIRs) of probabilistic amplitude shaping (PAS) for discrete-input memoryless channels.
In contrast to the existing literature in which Gallager's error exponent approach was followed, we used a weak typicality framework.
Random sign-coding arguments based on weak typicality were introduced to upper-bound the probability of error of a so-called sign-coding structure.
The achievability of the mutual information was demonstrated for uniform signs, which were independent of the amplitudes. 
Sign-coding combined with amplitude shaping corresponded to PAS, and~consequently, PAS achieved the capacity of a discrete-input memoryless channel with a symmetric capacity-achieving~distribution.

Our approach was different than the random coding arguments considered in the literature, in~the sense that our motivation was to provide achievability proofs that were as constructive as possible.
To~this end, in~our random sign-coding setup, both the amplitudes and the signs of the channel inputs that were directly selected by information bits were constructively produced.
Only the remaining signs were drawn at random.
A study on the achievability of capacity for channels with asymmetric capacity-achieving distributions with a type of sign-coding is left for possible future~research.

%
%
%
%
%

\section*{Appendix A \\ Proof of Lemma~\ref{Lemma.Ps}}\label{App.P2}

\subsection{Proof of~\ref{prop:weaktyp1}}
We see from~\cite[Eq. (3.6)]{CoverT2006_ElementsofInfoTheo} that for $\uu\in\Aeps(U)$,
\begin{equation}
 2^{-n(H(U)+\varepsilon)} \leq p(\uu) \leq 2^{-n(H(U)-\varepsilon)}. \label{eq:P1proof}
 \end{equation}
 Due to Definition~\ref{def:Beps}, each $\uu \in \BepsV(U)$ is also in $\Aeps(U)$; more specifically, $\BepsV{}(U) \subseteq \Aeps(U)$.
 Consequently, \eqref{eq:P1proof} also holds for $\uu\in\BepsV(U)$, which completes the proof of~\ref{prop:weaktyp1}.
 
\subsection{Proof of~\ref{prop:weaktyp2}}
Let $(\uU,\uV)$ be independent and identically distributed with respect to $p(u,v)$.
 Then:
 \begin{IEEEeqnarray}{rCl} 
\Pr\{(\uU,\uV) \in \Aeps(UV)\} 
 &=& \sum_{\uu} p(\uu) \sum_{\uv: (\uu,\uv) \in \Aeps(UV) } p(\uv|\uu) \\
 &=& \sum_{\uu\in\BepsV(U)} p(\uu) \sum_{\uv: (\uu,\uv) \in \Aeps(UV) } p(\uv|\uu) \nonumber \\
&& +\> \sum_{\uu\notin\BepsV(U)} p(\uu) \sum_{\uv: (\uu,\uv) \in \Aeps(UV)} p(\uv|\uu) \\
 &\leq& \sum_{\uu\in\BepsV(U)} p(\uu) + \sum_{\uu\notin\BepsV(U)} p(\uu) (1-\varepsilon) \label{p2r6} \\
 &=& 1- \varepsilon + \varepsilon \sum_{\uu\in\BepsV(U)} p(\uu) \\
 &=& 1- \varepsilon + \varepsilon \Pr\{\uU \in \BepsV(U) \}. \label{p2r9}
 \end{IEEEeqnarray}
 Here, \eqref{p2r6} follows from Definition~\ref{def:Beps}, which states that $\Pr\left\{(\uu,\uV)\in\Aeps(UV) \big| \uU = \uu\right\} < 1-\varepsilon$ for $\uu\in\Aeps(U)$, if~$\uu\notin\BepsV(U)$.
 Then, from~\eqref{p2r9}, we obtain:
 \begin{IEEEeqnarray}{rCl}
 \Pr\{\uU \in \BepsV(U) \} 
 &\geq& \frac{ \Pr\{(U,V) \in \Aeps(UV)\} -1 + \varepsilon}{\varepsilon} \\
 &=& 1 - \frac{\Pr\{(U,V) \notin \Aeps(UV) \} }{\varepsilon} \\
 &\geq& 1- \varepsilon. \label{p2r11}
 \end{IEEEeqnarray}
 for large enough $n$.
 Here, \eqref{p2r11} follows from~\cite[Thm. 7.6.1]{CoverT2006_ElementsofInfoTheo}, which states that $\Pr\{(\uU,\uV)\in\Aeps(UV)\}\rightarrow 1$ as $n\rightarrow\infty$.
 This implies that $\Pr \{ (\uU,\uV) \notin\Aeps(UV) \} \leq \varepsilon^2$ for positive $\varepsilon$ and large enough $n$, which completes the~proof.

\subsection{Proof of~\ref{prop:weaktyp3}}
We see from~\cite[Thm. 3.1.2]{CoverT2006_ElementsofInfoTheo} that:
\begin{equation}
|\Aeps(U)| \leq 2^{n(H(U)+\varepsilon)}. \label{eq:P3proof1}
\end{equation}
Since $\BepsV{}(U)\subseteq \Aeps(U)$, again by Definition~\ref{def:Beps}, \eqref{eq:P3proof1} also holds for $|\BepsV{}(U)|$.
This proves the upper bound in~\ref{prop:weaktyp3}.
To prove the lower bound, we obtain from \eqref{p2r11} for $n$ sufficiently large that:
\begin{IEEEeqnarray}{rCl} 
1-\varepsilon &\leq& \Pr\{\uU \in \BepsV(U) \} \\
&\leq& \sum_{\uu\in\BepsV(U)} 2^{-n(H(U)-\varepsilon)} \label{p1rw} \\
&=& |\BepsV(U)|2^{-n(H(U)-\varepsilon)},
\end{IEEEeqnarray}
where \eqref{p1rw} follows from \eqref{eq:P1proof}.

\section*{Appendix B \\ Proofs of Theorems~\ref{bsc_smd},~\ref{mod_smd},~\ref{bsc_bmd}, and~\ref{mod_bmd}}\label{sec:app_thms}
To derive AIRs, we will follow the classical approach, e.g.,~as in~\cite[Sec. 7.7]{CoverT2006_ElementsofInfoTheo}, and~upper-bound the average of the probability of error $\oPe$ over a random choice of sign-codebooks.
This way, we will demonstrate the existence of at least one good sign-code.
Again as in~\cite[Sec. 7.7]{CoverT2006_ElementsofInfoTheo} and as explained in Section~\ref{sec:decoding}, we decode by joint typicality: the decoder looks for a unique message index pair $(\hat{m}_a,\hat{m}_s)$ for which the corresponding amplitude-sign sequence $(\ua,\us)$ is jointly typical with the received sequence $\uy$.

By the properties of weak typicality and $\calB$-typicality, the~transmitted amplitude-sign sequence and the received sequence are jointly typical with high probability for $n$ large enough.
We call the event for which the transmitted amplitude-sign sequence is not jointly typical with the received sequence the {first error event} with average probability $\oPe(1)$.
Furthermore, the~probability that any other (not transmitted) amplitude-sign sequence is jointly typical with the received sequence vanishes for asymptotically large $n$.
We call the event that there is another amplitude-sign sequence that is jointly typical with the received sequence the {second error event} with average probability $\oPe(2)$.
Observing~that these events are {not} disjoint, we can write~\cite[Eq. (7.75)]{CoverT2006_ElementsofInfoTheo}:
\begin{equation}
 \oPe \leq \oPe(1) + \oPe(2). \label{appBdef}
\end{equation}

\subsection{Proof of Theorem~\ref{bsc_smd}}

For the error of the first kind, we can write:
\begin{IEEEeqnarray}{rCl} 
\oPe(1) &=& \sum_{m_a=1}^{M_a} \frac{1}{M_a} \sum_{\us\in\calS^n} p(\us) \sum_{\uy\in\calY^n} p(\uy|\ua(m_a),\us) \mathds{1}[(\ua(m_a),\us,\uy) \notin \Aeps(ASY)]\\
&=&\sum_{m_a} \frac{1}{M_a} \sum_{\us} \sum_{\uy} p(\us,\uy|\ua(m_a)) \mathds{1}[ (\ua(m_a),\us,\uy) \notin \Aeps] \label{t1p1r2} \\
&=& \sum_{m_a} \frac{1}{M_a} \Pr\left\{(\ua(m_a),\uS,\uY) \notin \Aeps \big| \uA=\ua(m_a) \right\} \\
&\leq& \sum_{m_a} \frac{\varepsilon}{M_a} \label{t1p1r4} \\
&=& \varepsilon, \label{eq:thm1pe1}
\end{IEEEeqnarray}
where we simplified the notation by replacing $m_a = 1, 2, \cdots, M_a$ by $m_a$, $\us\in\calS^n$ by $\us$, and~$\uy\in\calY^n$ by $\uy$ in \eqref{t1p1r2}.
Furthermore, we dropped the index of the typical set $\Aeps(ASY)$ and used $\Aeps$ instead.
We will follow these notations for summations and for the typical sets for the rest of the paper, assuming for the latter that the index of the typical set will be clear from the context.
To obtain \eqref{t1p1r2}, we~used $p(\us) p(\uy|\ua(m_a),\us) = p(\us,\uy|\ua(m_a))$.
Then, \eqref{t1p1r4} is a direct consequence of Definition \ref{def:Beps} since $\ua(m_a) \in \BepsSY(A)$ for $m_a = 1, 2,\cdots, M_a$.

For the error of the second kind, we can write:
\begin{IEEEeqnarray}{rCl} 
\oPe(2) &\leq& \sum_{m_a} \frac{1}{M_a} \sum_{\us} p(\us) \sum_{\uy} p(\uy|\ua(m_a),\us) \sum_{k_a=1, k_a \neq m_a}^{M_a} \sum_{\usp\in\calS^n} p(\usp) \mathds{1}[ (\ua(k_a),\usp,\uy) \in \Aeps ] \\
&=& M_a \sum_{m_a} \sum_{\us} \frac{p(\us)}{M_a} \sum_{\uy} p(\uy|\ua(m_a),\us) \sum_{k_a\neq m_a} \sum_{\usp} \frac{p(\usp)}{M_a} \mathds{1}[ (\ua(k_a),\usp,\uy) \in \Aeps ] \label{t1p2r3} \\
&\leq& M_a 2^{6n\varepsilon } \sum_{m_a} \sum_{\us} p(\ua(m_a)) p(\us) \sum_{\uy} p(\uy|\ua(m_a),\us) \nonumber \\
&& \cdot \sum_{k_a\neq m_a} \sum_{\usp} p(\ua(k_a)) p(\usp) \mathds{1}[ (\ua(k_a),\usp,\uy) \in \Aeps ] \label{t1p2r4} \\
&\leq& M_a 2^{6n\varepsilon } \sum_{\ua\in\calA^n} \sum_{\us} p(\ua)p(\us) \sum_{\uy} p(\uy|\ua,\us) \sum_{\uap\in\calA^n} \sum_{\usp} p(\uap)p(\usp) \mathds{1}[ (\uap,\usp,\uy) \in \Aeps ] \label{t1p2r5} \\
&=& M_a 2^{6n\varepsilon } \sum_{\substack{(\uy,\uxp) \in \Aeps} } p(\uxp)p(\uy) \label{t1p2r7} \\
&\leq& 2^{n(H(A)+\varepsilon)}2^{6n\varepsilon} |\Aeps(XY)| 2^{-n(H(X)-\varepsilon)}2^{-n(H(Y)-\varepsilon)} \label{t1p2end} \\
&\leq& 2^{n(H(A)+7\varepsilon)} 2^{n(H(X,Y)+\varepsilon)} 2^{-n(H(X)-\varepsilon)} 2^{-n(H(Y)-\varepsilon)} \label{t1p2end1} \\
&=& 2^{n(H(A)-I(SA;Y)+10\varepsilon)}, \label{t1p2end2}
\end{IEEEeqnarray}
where we simplified the notation by replacing $k_a = 1, 2,\cdots, M_a: k_a\neq m_a$ by $k_a\neq m_a$, and~$\usp\in\calS^n$ by $\usp$ in \eqref{t1p2r3}.
We will follow these notations for the rest of the paper.
Then:
\begin{itemize}
 \item[\eqref{t1p2r4}] follows for $n$ sufficiently large and for $\ua \in \BepsSY(A)$ from: 
 \begin{IEEEeqnarray}{rCl} 
 \frac{1}{M_a} = \frac{1}{|\BepsSY(A)|} &\leq& \frac{ 2^{-n(H(A)-\varepsilon))}}{1-\varepsilon} \label{crucial1}\\
 &=& \frac{ 2^{2n\varepsilon } }{1-\varepsilon} 2^{-n(H(A)+\varepsilon)} \\
 &\leq& \frac{ 2^{2n\varepsilon } }{1-\varepsilon} p(\ua) \label{crucial2} \\
 &\leq& 2^{3n\varepsilon } p(\ua), \label{crucial}
 \end{IEEEeqnarray}
 where \eqref{crucial1} follows from the $\calB$-typicality property \ref{prop:weaktyp3}, \eqref{crucial2} follows from the $\calB$-typicality property \ref{prop:weaktyp1}, and~\eqref{crucial} holds for all large enough $n$.
 
 \item[\eqref{t1p2r5}] follows from summing over $\ua\in\calA^n$ instead of over $\ua(m_a)\in\Beps$ and over $\uap\in\calA^n$ instead of $\ua(k_a)\in\Beps$ for $k_a\neq m_a$.
 
 \item[\eqref{t1p2r7}] is obtained by working out the summations over $\ua$ and $\us$ and by replacing $\uap\usp$ with $\uxp$.
 
 \item[\eqref{t1p2end}] follows from $M_a = |\Beps(A)| \leq 2^{n(H(A)+\varepsilon)}$, i.e.,~the $\calB$-typicality property \ref{prop:weaktyp3}, and~from \eqref{eq:typprob}.
 
 \item[\eqref{t1p2end1}] follows from \eqref{eq:jointtypcardi}.
\end{itemize}
The conclusion from \eqref{t1p2end2} is that for $H(A) < I(X;Y) -10\varepsilon$, the~error probability of the second kind:
\begin{equation}
\oPe(2) \leq \varepsilon \label{eq:thm1pe2}
\end{equation}
for $n$ large enough.
Using \eqref{eq:thm1pe1} and \eqref{eq:thm1pe2} in \eqref{appBdef}, we find that the total error probability averaged over all possible sign-codes $\oPe \leq 2\varepsilon$ for $n$ large enough. 
This implies the existence of a basic sign-code with total error probability $\Pe =\Pr\{\hat{M_a} \neq M_a\} \leq 2\varepsilon$. 
This holds for all $\varepsilon>0$, and~therefore, the~rate:
\begin{equation}
 R = H(A) \leq I(X;Y),
\end{equation}
is achievable with basic sign-coding, which concludes the proof of Theorem~\ref{bsc_smd}.

\subsection{Proof of Theorem~\ref{mod_smd}}

For the error of the first kind, we can write:
\begin{IEEEeqnarray}{rCl} 
\oPe(1) &=& \sum_{m_a} \frac{1}{M_a} \sum_{m_s=1}^{M_s} \frac{1}{2^{n_1}} \sum_{\ust\in\calS^{n_2}} p(\ust) \sum_{\uy} p(\uy|\ua(m_a),\use(m_s)\ust) \mathds{1}[(\ua(m_a),\use(m_s)\ust,\uy) \notin \Aeps] \label{t2p1r1} \\
&=& \sum_{m_a} \frac{1}{M_a} \sum_{m_s} \sum_{\ust} 2^{-n} \sum_{\uy} p(\uy|\ua(m_a),\use(m_s)\ust) \mathds{1}[(\ua(m_a),\use(m_s)\ust,\uy) \notin \Aeps] \label{t2p1r2} \\
&=& \sum_{m_a} \frac{1}{M_a} \sum_{m_s} \sum_{\ust} \sum_{\uy} p(\use(m_s)\ust,\uy|\ua(m_a)) \mathds{1}[(\ua(m_a),\use(m_s)\ust,\uy) \notin \Aeps] \label{t2p1r25} \\
&=& \sum_{m_a} \frac{1}{M_a} \Pr \left\{ (\ua(m_a),\uS,\uY) \notin \Aeps \big| \uA=\ua(m_a)\right\} \label{t2p1r3} \\
&\leq& \sum_{m_a} \frac{\varepsilon}{M_a} \label{t2p1r4} \\
&=& \varepsilon, \label{eq:thm2pe1}
\end{IEEEeqnarray}
where we simplified the notation by replacing $\ust\in\calS^{n_2}$ by $\ust$ and $m_s = 1, 2,\cdots, M_s$ by $m_s$ in \eqref{t2p1r2}.
We~will follow these notations for the rest of the paper.
To obtain \eqref{t2p1r2}, we used the fact that $\uSt$ is uniform; more precisely $p(\ust)=2^{-n_2}$.
To obtain \eqref{t2p1r25}, we used the fact that $\uSe$ is also uniform, and~then, $2^{-n}p(\uy|\ua(m_a),\use(m_s)\ust) = p(\use(m_s)\ust,\uy|\ua(m_a))$.
Then, \eqref{t2p1r4} is a direct consequence of Definition \ref{def:Beps} since $\ua(m_a) \in \BepsSY(A)$ for $m_a = 1, 2,\cdots, M_a$.

For the error of the second kind, we obtain:
\begin{IEEEeqnarray}{rCl} 
\oPe(2) &\leq& \sum_{m_a} \frac{1}{M_a} \sum_{m_s} \frac{1}{2^{n_1}} \sum_{\ust} p(\ust) \sum_{\uy} p(\uy|\ua(m_a),\use(m_s)\ust) \nonumber \\
&& \cdot \sum_{(k_a,k_s) \neq (m_a,m_s)} \sum_{\ustp} p(\ustp) \mathds{1}[ (\ua(k_a),\use(k_s)\ustp,\uy) \in \Aeps ] \nonumber \\
&=& M_a2^{n_1} \sum_{m_a,m_s,\ust} \frac{2^{-n}}{M_a} \sum_{\uy} p(\uy|\ua(m_a),\use(m_s)\ust) \nonumber \\
&& \cdot \sum_{(k_a,k_s) \neq (m_a,m_s)} \sum_{\ustp} \frac{2^{-n}}{M_a} \mathds{1}[ (\ua(k_a),\use(k_s)\ustp,\uy) \in \Aeps ] \label{t2p2r1} \\
&=& M_a 2^{n_1}\sum_{m_a,m_s,\ust} \frac{2^{-n}}{M_a} \sum_{\uy} p(\uy|\ua(m_a),\use(m_s)\ust) \sum_{k_a \neq m_a, k_s ,\ustp} \frac{2^{-n}}{M_a} \mathds{1}[ (\ua(k_a),\use(k_s)\ustp,\uy) \in \Aeps ] \nonumber \\
&& +\> 2^{n_1}\sum_{m_a,m_s,\ust} \frac{2^{-n}}{M_a} \sum_{\uy} p(\uy|\ua(m_a),\use(m_s)\ust) \sum_{k_s \neq m_s ,\ustp} 2^{-n} \mathds{1}[ (\ua(m_a),\use(k_s)\ustp,\uy) \in \Aeps ]. \label{t2p2r2} 
\end{IEEEeqnarray}
Here, we replaced nested summations over $m_a$, $m_s$, and~$\use$ by a single summation over $(m_a,m_s,\use)$ for the sake of better readability.
We will use this notation for the rest of the paper.
Then:
\begin{itemize}
 \item[\eqref{t2p2r1}] follows from $n = n_1+n_2$ and from the fact that $\uSt$ is uniform; more precisely, $p(\ust)=2^{-n_2}$.
 
 \item[\eqref{t2p2r2}] is obtained by splitting $(k_a,k_s) \neq (m_a,m_s)$ into $k_a\neq m_a, k_s$ and $k_a=m_a, k_s\neq m_s$.
\end{itemize}
From \eqref{t2p2r2}, we obtain:
\begin{IEEEeqnarray}{rCl} 
\oPe(2) &\leq& M_a 2^{n_1} 2^{6n\varepsilon} \sum_{m_a,m_s,\ust} p(\ua(m_a)) p(\use(m_s)\ust) \sum_{\uy} p(\uy|\ua(m_a),\use(m_s)\ust) \nonumber \\
&& \IEEEeqnarraynumspace \cdot \sum_{k_a\neq m_a, k_s,\ustp} p(\ua(k_a)) p(\use(k_s)\ustp)\mathds{1}[ (\ua(k_a),\use(k_s)\ustp,\uy) \in \Aeps] \nonumber \\
&& +\> 2^{n_1} 2^{3n\varepsilon} \sum_{m_a,m_s,\ust} p(\ua(m_a)) p(\use(m_s)\ust) \sum_{\uy} p(\uy|\ua(m_a),\use(m_s)\ust) \nonumber \\
&& \IEEEeqnarraynumspace \cdot \sum_{k_s \neq m_s,\ustp} p(\use(k_s)\ustp) \mathds{1}[ (\ua(m_a),\use(k_s)\ustp,\uy) \in \Aeps ] \label{t2p2r3} \\
&\leq& M_a 2^{n_1} 2^{6n\varepsilon} \sum_{\ua, \use\ust} p(\ua) p(\use\ust) \sum_{\uy} p(\uy|\ua,\use\ust) \sum_{\uap, \usep \ustp} p(\uap) p(\usep \ustp) \mathds{1}[ (\uap,\usep\ustp,\uy) \in \Aeps ] \nonumber \\
&& +\> 2^{n_1} 2^{3n\varepsilon} \sum_{\ua, \use\ust} p(\ua) p(\use\ust) 
 \sum_{\uy} p(\uy|\ua,\use\ust) \sum_{\usep\ustp} p(\usep\ustp) 
 \mathds{1}[ (\ua,\usep\ustp,\uy) \in \Aeps ] \label{t2p2r4} \\ 
&=& M_a 2^{n_1} 2^{6n\varepsilon} \sum_{\ua,\us} p(\ua)p(\us) \sum_{\uy} p(\uy|\ua,\us) \sum_{\uap,\usp} p(\uap)p(\usp) \mathds{1}[(\uap,\usp,\uy) \in \Aeps] \nonumber \\
&& +\> 2^{n_1} 2^{3n\varepsilon} \sum_{\ua,\us} p(\ua) p(\us) \sum_{\uy} p(\uy|\ua,\us) \sum_{\usp} p(\usp) \mathds{1}[(\ua,\usp,\uy) \in \Aeps], \label{t2p2r5} 
\end{IEEEeqnarray}
where:
\begin{itemize}
 \item[\eqref{t2p2r3}] follows for $n$ sufficiently large and for $\ua \in \BepsSY(A)$ from:
\begin{equation}
 \frac{1}{M_a} \stackrel{\eqref{crucial}}{\leq} 2^{3n\varepsilon } p(\ua)
 \end{equation}
 and from $p(\use \ust)=2^{-n}$,
 
 \item[\eqref{t2p2r4}] follows from summing over $\ua\in\calA^n$ instead of over $\ua(m_a)\in\Beps$ and over $\uap\in\calA^n$ instead of $\ua(k_a)\in\Beps$ for $k_a\neq m_a$.
 Moreover, it follows from summing over $\use\in\calS^{n_1}$ instead of $\use(k_s)$ for $k_s = 1, 2,\cdots, M_s$ and $k_s\neq m_s$.
 
 \item[\eqref{t2p2r5}] follows from substituting $\us$ for $\use \ust$ and $\usp$ for $\usep \ustp$.
\end{itemize}
Finally, from~\eqref{t2p2r5}, we obtain:
\begin{IEEEeqnarray}{rCl} 
\oPe(2) &=& M_a 2^{n_1} 2^{6n\varepsilon} \sum_{\uy} p(\uy) \sum_{\uxp} p(\uxp) \mathds{1}[(\uxp,\uy) \in \Aeps] \nonumber \\
&& +\> 2^{n_1} 2^{3n\varepsilon} \sum_{\ua,\uy} p(\ua,\uy) \sum_{\usp} p(\usp) \mathds{1}[(\ua,\usp,\uy) \in \Aeps] \label{t2p2r6} \\
&\leq& 2^{n(H(A)+\varepsilon)} 2^{n\gamma} 2^{6n\varepsilon} |\Aeps(XY)| 2^{-n(H(X)-\varepsilon)}2^{-n(H(Y)-\varepsilon)} \nonumber \\
&& +\> 2^{n\gamma} 2^{3n\varepsilon} |\Aeps(SAY)| 2^{-n(H(A,Y)-\varepsilon)} 2^{-n(H(S)-\varepsilon)} \label{t2p2r7} \\
&\leq& 2^{n(H(A)+7\varepsilon)} 2^{n\gamma} 2^{n(H(X,Y)+\varepsilon)} 2^{-n(H(X)-\varepsilon)}2^{-n(H(Y)-\varepsilon)} \nonumber \\
&& +\> 2^{n\gamma} 2^{3n\varepsilon} 2^{n(H(S,A,Y)+\varepsilon)} 2^{-n(H(A,Y)-\varepsilon)} 2^{-n(H(S)-\varepsilon)} \label{t2p2r8} \\
&=& 2^{n(H(A)+\gamma+10\varepsilon-I(X;Y))} + 2^{n(\gamma+6\varepsilon-I(S;A,Y))}. \label{t2p2r9}
\end{IEEEeqnarray}
Here, we substituted $n_1 = n \gamma$ in \eqref{t2p2r7}.
Then:
\begin{itemize}
 \item[\eqref{t2p2r6}] is obtained by working out the summations over $\ua,\us$ in the first part and $\us$ in the second part.
 Moreover, we replaced $\uap\usp$ with $\uxp$.
 
 \item[\eqref{t2p2r7}] is obtained using for the first part that $M_a = |\Beps(A)| \leq 2^{n(H(A)+\varepsilon)}$, i.e.,~the $\calB$-typicality property \ref{prop:weaktyp3}, and~\eqref{eq:typprob}.
 For the second part, we used \eqref{eq:typprob} for $p(\us)$ and \eqref{eq:jointtypprob} for $p(\ua,\uy)$.
 
 \item[\eqref{t2p2r8}] follows from \eqref{eq:jointtypcardi}, and~its extension to jointly typical triplets; more precisely, \mbox{$|\Aeps(SAY)| \leq 2^{n(H(S,A,Y)+\varepsilon)}$}.
 
\end{itemize}

The conclusion from \eqref{t2p2r9} is that for $H(A)+\gamma < I(X;Y) - 10\varepsilon$ and $\gamma < I(S;A,Y) - 6\varepsilon$, the~error probability of the second kind:
\begin{equation}
\oPe(2) \leq \varepsilon, \label{eq:thm2pe2}
\end{equation} 
for $n$ large enough.
The first constraint, i.e.,~$H(A)+\gamma < I(X;Y) - 10\varepsilon$, already implies the second constraint, i.e.,~$\gamma < I(S;A,Y) - 6\varepsilon$, since:
\begin{IEEEeqnarray}{rCl} 
\gamma &<& I(X;Y) - H(A) - 10\varepsilon \nonumber\\
&\leq& I(S,A;Y) - I(A;Y)- 10\varepsilon \label{eq:2ndr2} \\
&=& I(S;Y|A)- 10\varepsilon \label{eq:2ndr3} \\
&\leq& I(S;Y|A) + I(S;A) - 10\varepsilon\\
&=& I(S;A,Y)- 10\varepsilon, \label{eq:2ndr5}
\end{IEEEeqnarray}
where we substituted $(S,A)$ for $X$ in \eqref{eq:2ndr2}.
Here, \eqref{eq:2ndr2} follows from~\cite[Thm. 2.4.1]{CoverT2006_ElementsofInfoTheo}, and~both \eqref{eq:2ndr3} and \eqref{eq:2ndr5} follow from the chain rule for MI~\cite[Thm. 2.5.2]{CoverT2006_ElementsofInfoTheo}.

Using \eqref{eq:thm2pe1} and \eqref{eq:thm2pe2} in \eqref{appBdef}, we find that the total error probability averaged over all possible modified sign-codes $\oPe \leq 2\varepsilon$ for $n$ large enough.
This implies the existence of a modified sign-code with total error probability $\Pe = \Pr\{(\hat{M}_a,\hat{M}_s) \neq (M_a,M_s)\} \leq 2\varepsilon$.
This holds for all $\varepsilon>0$, and~thus, the~rate:
\begin{equation}
 R = H(A) +\gamma \leq I(X;Y),
\end{equation}
is achievable with modified sign-coding, which concludes the proof of Theorem~\ref{mod_smd}.

\subsection{Proof of Theorem~\ref{bsc_bmd}}
For the error of the first kind, we can write:
\begin{IEEEeqnarray}{rCl} 
\oPe(1) &=& \sum_{m_a} \frac{1}{M_a} \sum_{\us} p(\us) \sum_{\uy} p(\uy|\uboldb(m_a),\us) \label{t3p1r1} \\
&& \hfill \cdot \mathds{1}[((\ube(m_a),\uy) \notin \Aeps) \cup ((\ubt(m_a),\uy) \notin \Aeps) \cup \dotsc \cup ((\ub_\mm(m_a),\uy) \notin \Aeps) \cup ((\us,\uy) \notin \Aeps)] \nonumber \\
&\leq& \sum_{m_a} \frac{1}{M_a} \sum_{\us} \sum_{\uy} p(\us,\uy|\uboldb(m_a)) \mathds{1}[(\uboldb(m_a),\us,\uy) \notin \Aeps] \label{t3p1r2} \\
&=& \sum_{m_a} \frac{1}{M_a} \Pr\left\{(\uboldb(m_a),\uS,\uY) \notin \Aeps \big| \uboldB=\uboldb(m_a) \right\} \label{t3p1r3} \\
&\leq& \sum_{m_a} \frac{\varepsilon}{M_a} \label{t3p1r4} \\
&=& \varepsilon, \label{eq:thm3pe1}
\end{IEEEeqnarray}
where we used $\uboldb(m_a)$ to denote $(\ube(m_a),\ubt(m_a),\dotsc,\ub_\mm(m_a))$ in \eqref{t3p1r1} and $\uboldB$ to denote $(\uB_1,\uB_2,\dotsc,\uB_\mm)$ in \eqref{t3p1r3}.
Then, we used $p(\us)p(\uy|\uboldb(m_a),\us)=p(\us,\uy|\uboldb(m_a))$ in \eqref{t3p1r2}.
Here, \eqref{t3p1r2} follows from the fact that if at least one of $\ube(m_a), \ubt(m_a),\dotsc, \ub_\mm(m_a)$ or $\us$ is not jointly typical with $\uy$, then $(\uboldb(m_a),\us,\uy)$ is not jointly typical.
Then, \eqref{t3p1r4} is a direct consequence of Definition~\ref{def:Beps} since $\uboldb(m_a) \in \BepsSY(B_1 B_2 \cdots B_m)$ for $m_a = 1, 2,\cdots, M_a$.

For the error of the second kind, we can write:
\begin{IEEEeqnarray}{rCl} 
\oPe(2) &\leq& \sum_{m_a} \frac{1}{M_a} \sum_{\us} p(\us) \sum_{\uy} p(\uy|\uboldb(m_a),\us) \nonumber \\
&& \hfill \cdot \sum_{k_a\neq m_a} \sum_{\usp} p(\usp) 
 \mathds{1}[ (\ube(k_a),\uy) \in \Aeps, (\ubt(k_a),\uy) \in \Aeps, \dotsc, (\ub_\mm(k_a),\uy) \in \Aeps, (\usp,\uy) \in \Aeps ] \nonumber \\ 
&=& M_a \sum_{m_a}\sum_{\us} \frac{p(\us)}{M_a} \sum_{\uy} p(\uy|\uboldb(m_a),\us) \nonumber \\
&& \hfill \cdot \sum_{k_a \neq m_a} \sum_{\usp} \frac{p(\usp)}{M_a} \mathds{1}[(\ube(k_a),\uy) \in \Aeps, (\ubt(k_a),\uy) \in \Aeps, \dotsc, (\ub_\mm(k_a),\uy) \in \Aeps, (\usp,\uy) \in \Aeps] \nonumber \\
&\leq& M_a 2^{6n\varepsilon} \sum_{m_a}\sum_{\us} p(\uboldb(m_a)) p(\us) \sum_{\uy} p(\uy|\uboldb(m_a),\us) \label{t3p2r4} \\
&& \hfill \cdot \sum_{k_a \neq m_a} \sum_{\usp}p(\usp) 
 p(\uboldb(k_a)) \mathds{1}[(\ube(k_a),\uy) \in \Aeps, (\ubt(k_a),\uy) \in \Aeps, \dotsc, (\ub_\mm(k_a),\uy) \in \Aeps, (\usp,\uy) \in \Aeps] \nonumber\\
&\leq& M_a 2^{6n\varepsilon} \sum_{\uboldb \in \{0,1\}^{\mm n}} \sum_{\us} p(\uboldb)p(\us)\sum_{\uy} p(\uy|\uboldb,\us) \label{t3p2r5}\\
&& \hfill \cdot \sum_{\uboldbt\in\{0,1\}^{\mm n}} \sum_{\usp} p(\usp)p(\uboldbt) \mathds{1}[(\ubep,\uy) \in \Aeps, (\ubtp,\uy) \in \Aeps, \dotsc , (\underline{\tilde{b}}_\mm,\uy) \in \Aeps, (\usp,\uy) \in \Aeps] \nonumber \\
&=& M_a 2^{6n\varepsilon} \sum_{\uy} p(\uy) \sum_{\uboldbt,\usp} p(\uboldbt,\usp) \mathds{1}[(\ubep,\uy) \in \Aeps, (\ubtp,\uy) \in \Aeps, \dotsc, (\underline{\tilde{b}}_\mm,\uy) \in \Aeps, (\usp,\uy) \in \Aeps] \IEEEeqnarraynumspace \label{t3p2r6}\\
&\leq& 2^{n(H(\boldB)+7\varepsilon)} |\Aeps(Y)| 2^{-n(H(Y)-\varepsilon)} \nonumber \\
&& \hfill \cdot |\Aeps(B_1|\uy)| |\Aeps(B_2|\uy)| \cdot \dotsc \cdot |\Aeps(B_\mm|\uy)| |\Aeps(S|\uy)|2^{-n(H(\boldB,S)-\varepsilon)} \IEEEeqnarraynumspace \label{t3p2r7}\\
&\leq& 2^{n(H(\boldB)+7\varepsilon)} 2^{n(H(Y)+\varepsilon)} 2^{-n(H(Y)-\varepsilon)} \nonumber \\
&& \hfill \cdot 2^{n(H(B_1|Y)+H(B_2|Y)+\dotsc+H(B_\mm|Y)+H(S|Y)+2(\mm+1)\varepsilon)} 2^{-n(H(\boldB,S)-\varepsilon)} \IEEEeqnarraynumspace \label{t3p2r8} \\
&=& 2^{n(H(\boldB)-H(\boldB,S) + H(B_1|Y) + H(B_2|Y) + \dotsc + H(B_\mm|Y) + H(S|Y) + (12+2\mm)\varepsilon)}, \label{t3p2r9}
\end{IEEEeqnarray}
where we used $\uboldb$ to denote $(\ube,\ubt,\dotsc,\ub_\mm)$ and $\uboldbt$ to denote $(\ubep,\ubtp,\dotsc,\underline{\tilde{b}}_\mm)$ in \eqref{t3p2r5}.
We also used $\boldB$ to denote $(B_1,B_2,\dotsc,B_\mm)$ in \eqref{t3p2r7}.
Finally, we simplified the notation by replacing $\uboldbt\in\{0,1\}^{\mm n}$ by $\uboldbt$ in \eqref{t3p2r6}.
Then:
\begin{itemize}
 \item[\eqref{t3p2r4}] follows for $n$ sufficiently large and for $\uboldb \in \BepsSY(\boldB)$ from $1/M_a \leq 2^{3n\varepsilon}p(\uboldb)$, which can be shown in a similar way as \eqref{crucial} was derived.
 \item[\eqref{t3p2r5}] follows from summing over $\uboldb \in \{0,1\}^{\mm n}$ instead of over $\uboldb(m_a)\in\Beps$ and over $\uboldbt \in \{0,1\}^{\mm n}$ instead of over $\uboldb(k_a) \in \Beps$ for $k_a\neq m_a$.
 \item[\eqref{t3p2r6}] is obtained by working out the summations over $\ube,\ubt, \dotsc, \ub_\mm$, and~$\us$.
 \item[\eqref{t3p2r7}] follows from $M_a = |\Beps(\boldB)| \leq 2^{n(H(\boldB)+\varepsilon)}$, i.e.,~the $\calB$-typicality property~\ref{prop:weaktyp3}, from~\eqref{eq:typprob}, and~from \eqref{eq:3jointtypprob}.
 \item[\eqref{t3p2r8}] follows from \eqref{eq:typcardi} and \eqref{eq:condtypcardi}.
\end{itemize}
The conclusion from \eqref{t3p2r9} is that for: 
\begin{IEEEeqnarray}{rCl} 
H(\boldB)&<&H(\boldB,S)-H(S|Y) - \left( \sum_{i=1}^{\mm} H(B_i|Y) \right)- (12+2\mm) \varepsilon \nonumber \\
 &=& \rbmd(p(\boldb,s))- (12+2\mm) \varepsilon, \nonumber
\end{IEEEeqnarray}
the error probability of the second kind:
\begin{equation}
\oPe(2) \leq \varepsilon \label{eq:thm3pe2}
\end{equation} 
for $n$ large enough.
Using \eqref{eq:thm3pe1} and \eqref{eq:thm3pe2} in \eqref{appBdef}, we find that the total error probability averaged over all possible sign-codes $\oPe \leq 2\varepsilon$ for $n$ large enough.
This implies the existence of a sign-code with total error probability $\Pe = \Pr\{\hat{M}_a\neq M_a\} \leq 2\varepsilon$.
This holds for all $\varepsilon>0$, and~thus, the~rate:
\begin{equation}
 R = H(\boldB) \leq \rbmd
\end{equation}
is achievable with sign-coding and BMD, which concludes the proof of Theorem~\ref{bsc_bmd}.

\subsection{Proof of Theorem~\ref{mod_bmd}}
For the error of first kind, we can write:
\begin{IEEEeqnarray}{rCl} 
\oPe(1) &=& \sum_{m_a} \frac{1}{M_a} \sum_{m_s} \frac{1}{2^{n_1}} \sum_{\ust} p(\ust) \sum_{\uy} p(\uy|\uboldb(m_a),\use(m_s)\ust) \nonumber \\
&& \hfill \cdot \mathds{1}\left[ \bigcup_{i=1}^m ( (\ub_i(m_a),\uy)\notin\Aeps ) \bigcup ( (\use(m_s)\ust,\uy)\notin\Aeps ) \right] \nonumber \\
&=& \sum_{m_a} \frac{1}{M_a} \sum_{m_s} \sum_{\ust} 2^{-n} \sum_{\uy} p(\uy|\uboldb(m_a),\use(m_s)\ust) \nonumber \\
&& \hfill \cdot \mathds{1}\left[ \bigcup_{i=1}^m ( (\ub_i(m_a),\uy)\notin\Aeps ) \bigcup ( (\use(m_s)\ust,\uy)\notin\Aeps ) \right] \label{t4p1r2}\\
&\leq& \sum_{m_a} \frac{1}{M_a} \sum_{m_s} \sum_{\ust} \sum_{\uy} p(\use(m_s)\ust,\uy|\uboldb(m_a)) \mathds{1}[(\uboldb(m_a),\use(m_s)\ust,\uy)\notin\Aeps] \label{t4p1r3}\\
&=& \sum_{m_a} \frac{1}{M_a} \Pr\{(\uboldb(m_a),\uS,\uY)\notin\Aeps|\uboldB=\uboldb(m_a)\} \nonumber\\
&\leq& \sum_{m_a} \frac{\varepsilon}{M_a} \label{t4p1r5}\\
&=& \varepsilon. \label{eq:thm4pe1}
\end{IEEEeqnarray}
Here, to~obtain~\eqref{t4p1r2}, we used the fact that $\uSt$ is uniform; more precisely, $p(\ust)=2^{-n_2}$.
Then, we~used $2^{-n}p(\uy|\uboldb(m_a),\use(m_s)\ust) = p(\use(m_s)\ust,\uy|\uboldb(m_a))$ in \eqref{t4p1r3}.
Furthermore, \eqref{t4p1r3} also follows from the fact that if at least one of $\ube(m_a), \ubt(m_a),\dotsc, \ub_\mm(m_a)$ or $\use(m_s)\ust$ is not jointly typical with $\uy$, \mbox{then $(\uboldb(m_a),\use(m_s)\ust,\uy)$} is not jointly typical.
Then, \eqref{t4p1r5} is a direct consequence of Definition~\ref{def:Beps} since $\uboldb(m_a) \in \BepsSY(B_1 B_2 \cdots B_m)$ for $m_a = 1, 2, \cdots, M_a$.

For the error of second kind, we can write:
\begin{IEEEeqnarray}{rCl} 
\oPe(2)
&\leq& \sum_{m_a} \frac{1}{M_a} \sum_{m_s} \frac{1}{2^{n_1}} \sum_{\ust} p(\ust) \sum_{\uy} p(\uy|\uboldb(m_a),\use(m_s)\ust) \nonumber \\
&& \hfill \cdot \sum_{(k_a,k_s) \neq (m_a,m_s)} \sum_{\ustp} p(\ustp) 
 \mathds{1} \left[ \bigcap_{i=1}^m ((\ub_i(k_a),\uy)\in\Aeps) \bigcap ((\use(k_s)\ustp,\uy)\in\Aeps) \right] \nonumber \\
&=& M_a2^{n_1} \sum_{m_a,m_s,\ust} \frac{2^{-n}}{M_a} \sum_{\uy} p(\uy|\uboldb(m_a),\use(m_s)\ust) \nonumber \\
&& \hfill \cdot \sum_{(k_a,k_s) \neq (m_a,m_s)} \sum_{\ustp} \frac{2^{-n}}{M_a} \mathds{1} \left[ \bigcap_{i=1}^m ((\ub_i(k_a),\uy)\in\Aeps) \bigcap ((\use(k_s)\ustp,\uy)\in\Aeps) \right] \label{t4p2r1} \\
&=& M_a 2^{n_1} \sum_{m_a, m_s, \ust} \frac{2^{-n}}{M_a} \sum_{\uy} p(\uy|\uboldb(m_a),\use(m_s)\ust) \nonumber \\
&& \hfill \cdot \sum_{k_a\neq m_a, k_s,\ustp} \frac{2^{-n}}{M_a} \mathds{1} \left[ \bigcap_{i=1}^m ((\ub_i(k_a),\uy)\in\Aeps) \bigcap ((\use(k_s)\ustp,\uy)\in\Aeps) \right] \nonumber \\
&& +\> 2^{n_1} \sum_{m_a,m_s,\ust} \frac{2^{-n}}{M_a} \sum_{\uy} p(\uy|\uboldb(m_a),\use(m_s)\ust) \nonumber \\
&& \hfill \cdot \sum_{k_s\neq m_s, \ustp} 2^{-n} \mathds{1} \left[ \bigcap_{i=1}^m ((\ub_i(m_a),\uy)\in\Aeps) \bigcap ((\use(k_s)\ustp,\uy)\in\Aeps) \right], \label{t4p2r2} 
\end{IEEEeqnarray}
where \eqref{t4p2r1} follows from $n=n_1+n_2$ and from the fact that $\uSt$ is uniform; more precisely, \mbox{$p(\ust)=2^{-n_2}$}.
Then, \eqref{t4p2r2} is obtained by splitting $(k_a,k_s) \neq (m_s,m_s)$ into $k_a\neq m_s, k_s$ and $k_a = m_a, k_s\neq m_s$.

From \eqref{t4p2r2}, we obtain:
\begin{IEEEeqnarray}{rCl} 
\oPe(2) &\leq& M_a 2^{n_1} 2^{6n\varepsilon} \sum_{m_a,m_s,\ust} p(\uboldb(m_a)) p(\use(m_s)\ust) \sum_{\uy} p(\uy|\uboldb(m_a),\use(m_s)\ust) \nonumber \\
&& \hfill \cdot \sum_{k_a\neq m_a, k_s, \ustp} p(\uboldb(k_a)) p(\use(k_s)\ustp) \mathds{1} \left[ \bigcap_{i=1}^m ((\ub_i(k_a),\uy)\in\Aeps) \bigcap ((\use(k_s)\ustp,\uy)\in\Aeps) \right] \nonumber \\
&& +\> 2^{n_1} 2^{3n\varepsilon} \sum_{m_a,m_s,\ust} p(\uboldb(m_a)) p(\use(m_s)\ust) \sum_{\uy} p(\uy|\uboldb(m_a),\use(m_s)\ust) \nonumber \\
&& \hfill \cdot \sum_{k_s\neq m_s, \ustp} p(\use(k_s)\ustp) \mathds{1} \left[ \bigcap_{i=1}^m ((\ub_i(m_a),\uy)\in\Aeps) \bigcap ((\use(k_s)\ustp,\uy)\in\Aeps) \right] \IEEEeqnarraynumspace \label{t4p2r3} \\
&\leq& M_a 2^{n_1} 2^{6n\varepsilon} \sum_{\uboldb,\use\ust} p(\uboldb)p(\use\ust) \sum_{\uy} p(\uy|\uboldb,\use\ust) \sum_{\uboldbt,\usep\ustp} p(\uboldbt) p(\usep\ustp) \nonumber \\ 
&& \hfill \cdot \mathds{1} \left[ \bigcap_{i=1}^m ((\ubtt_i,\uy)\in\Aeps) \bigcap ((\usep\ustp,\uy)\in\Aeps) \right] \IEEEeqnarraynumspace \nonumber \\ 
&& +\> 2^{n_1} 2^{3n\varepsilon} \sum_{\uboldb,\use\ust} p(\uboldb) p(\use\ust) \sum_{\uy} p(\uy|\uboldb,\use\ust) \sum_{\usep\ustp} p(\usep\ustp) \nonumber \\
&& \hfill \cdot \mathds{1} \left[ \bigcap_{i=1}^m ((\ub_i,\uy)\in\Aeps) \bigcap ((\usep\ustp,\uy)\in\Aeps) \right] \IEEEeqnarraynumspace \label{t4p2r4} \\
&=& M_a 2^{n_1} 2^{6n\varepsilon} \sum_{\uboldb,\us} p(\uboldb) p(\us) \sum_{\uy} p(\uy|\uboldb,\us) \sum_{\uboldbt,\usp} p(\uboldbt) p(\usp) \mathds{1} \left[ \bigcap_{i=1}^m ((\ubtt_i,\uy)\in\Aeps) \bigcap ((\usp,\uy)\in\Aeps) \right] \nonumber \\
&& +\> 2^{n_1} 2^{3n\varepsilon} \sum_{\uboldb,\us} p(\uboldb) p(\us) \sum_{\uy} p(\uy|\uboldb,\us) \sum_{\usp} p(\usp) \mathds{1} \left[ \bigcap_{i=1}^m ((\ub_i,\uy)\in\Aeps) \bigcap ((\usp,\uy)\in\Aeps) \right], \label{t4p2r5}
\end{IEEEeqnarray}
where:
\begin{itemize}
 \item[\eqref{t4p2r3}] follows for $n$ sufficiently large and for $\uboldb \in \BepsSY(\boldB)$ from $1/M_a \leq 2^{3n\varepsilon} p(\uboldb)$ and from \mbox{$p(\use\ust)=2^{-n}$},
 \item[\eqref{t4p2r4}] follows from summing over $\uboldb \in \{0, 1\}^{\mm n}$ instead of over $\uboldb(m_a) \in \Beps$ and over $\uboldbt \in \{0,1\}^{\mm n}$ instead of $\uboldb(k_a) \in \Beps$ for $k_a \neq m_a$. Moreover, it follows from summing over $\use \in \calS^{n_1}$ instead of $\use(k_s)$ for $k_s = 1, 2, \cdots, M_s$ and $k_s\neq m_s$,
 \item[\eqref{t4p2r5}] follows from substituting $\us$ for $\use\ust$ and $\usp$ for $\usep\ustp$.
\end{itemize}

Finally, from~\eqref{t4p2r5}, we obtain:
\begin{IEEEeqnarray}{rCl} 
\oPe(2) &=& M_a 2^{n_1} 2^{6n\varepsilon} \sum_{\uy} p(\uy) \sum_{\uboldbt,\usp} p(\uboldbt,\usp) \mathds{1} \left[ \bigcap_{i=1}^m ((\ubtt_i,\uy)\in\Aeps) \bigcap ((\usp,\uy)\in\Aeps) \right] \nonumber \\
&& +\> 2^{n_1} 2^{3n\varepsilon} \sum_{\uboldb,\uy} p(\uboldb,\uy) \sum_{\usp} p(\usp) \mathds{1} \left[ \bigcap_{i=1}^m ((\ub_i,\uy)\in\Aeps) \bigcap ((\usp,\uy)\in\Aeps) \right] \label{t4p2r6} \\
&\leq& 2^{n(H(\boldB)+\varepsilon)} 2^{n\gamma} 2^{6n\varepsilon} |\Aeps(Y)| 2^{-n(H(Y)-\varepsilon)} \left( \prod_{i=1}^m |\Aeps(B_i|\uy)| \right) |\Aeps(S|\uy)| 2^{-n(H(B_1B_2 \cdots B_m S)-\varepsilon)} \nonumber \\
&& +\> 2^{n\gamma} 2^{3n\varepsilon} |\Aeps(Y)| 2^{-n(H(\boldB Y)-\varepsilon)} 2^{-n(H(S)-\varepsilon)} \left( \prod_{i=1}^\mm |\Aeps(B_i|\uy)| \right) |\Aeps(S|\uy)| \label{t4p2r7} \\
&\leq& 2^{n(H(\boldB)+\varepsilon)} 2^{n\gamma} 2^{6n\varepsilon} 2^{n(H(Y)+\varepsilon)} 2^{-n(H(Y)-\varepsilon)} \left( \prod_{i=1}^\mm 2^{n(H(B_i|Y)+2\varepsilon)} \right) 2^{n(H(S|Y)+2\varepsilon)} 2^{-n(H(\boldB S)-\varepsilon)} \nonumber \\
&& +\> 2^{n\gamma} 2^{3n\varepsilon} 2^{n(H(Y)+\varepsilon)} 2^{-n(H(\boldB Y)-\varepsilon)} 2^{-n(H(S)-\varepsilon)} \left( \prod_{i=1}^\mm 2^{n(H(B_i|Y)+2\varepsilon)}\right) 2^{n(H(S|Y)+2\varepsilon)} \label{t4p2r8} \\
&=& 2^{n\left(H(\boldB)+\gamma+\left(\sum_{i=1}^\mm H(B_i|Y) \right) +H(S|Y)-H(\boldB S)+(12+2\mm)\varepsilon\right)} \nonumber \\
&& +\> 2^{n\left(\gamma+H(Y)-H(\boldB Y)-H(S)+\left(\sum_{i=1}^\mm H(B_i|Y) \right)+H(S|Y)+(8+2\mm)\varepsilon\right)}. \label{t4p2r9} 
\end{IEEEeqnarray}
Here, we substituted $n_1 = n\gamma$ in \eqref{t4p2r7}. Then:

\begin{itemize}
 \item[\eqref{t4p2r6}] is obtained by working out the summations over $\ube,\ubt,\dotsc,\ub_\mm,\us$ in the first part and $\us$ in the second part.
 \item[\eqref{t4p2r7}] is obtained using for the first part that $M_a = |\Beps(\boldB)| \leq 2^{n(H(\boldB)+\varepsilon)}$, i.e.,~the $\calB$-typicality property~\ref{prop:weaktyp3}, \eqref{eq:typprob} for $p(\uy)$, and~\eqref{eq:3jointtypprob} for $p(\uboldbt,\usp)$. For~the second part, we used \eqref{eq:typprob} for $p(\usp)$ and \eqref{eq:3jointtypprob} for $p(\uboldb,\uy)$.
 \item[\eqref{t4p2r8}] follows from \eqref{eq:typcardi} and \eqref{eq:condtypcardi}.
\end{itemize}
The conclusion from \eqref{t4p2r9} is that for:
\begin{equation}
H(\boldB)+\gamma \leq \rbmd-(12+2\mm)\varepsilon, \label{lastcond1}
\end{equation}
and for:
\begin{equation}
\gamma \leq H(\boldB Y)+H(S)-H(Y)-\left(\sum_{i=1}^\mm H(B_i|Y) \right)-H(S|Y)-(8+2\mm)\varepsilon, \label{lastcond}
\end{equation}
the error probability of the second kind:
\begin{equation}
\oPe(2) \leq \varepsilon \label{eq:thm4pe2}
\end{equation}
for $n$ large enough. 
The second constraint \eqref{lastcond} is already implied by the first constraint \eqref{lastcond1} since:
\begin{IEEEeqnarray}{rCl} 
\gamma &\leq& H(\boldB Y)+H(S)-H(Y)-\left(\sum_{i=1}^\mm H(B_i|Y)\right)-H(S|Y)-(8+2\mm)\varepsilon \\
&=& H(\boldB Y)+H(S)-H(Y)-\left(\sum_{i=1}^\mm H(B_i|Y)\right)-H(S|Y) + H(\boldB S)-H(\boldB S)-(8+2\mm)\varepsilon \\
&=& H(\boldB Y)+H(S)-H(Y)+\rbmd-H(\boldB)-H(S)-(8+2\mm)\varepsilon \\
&=& H(\boldB|Y) + \rbmd - H(\boldB) -(8+2\mm)\varepsilon.
\end{IEEEeqnarray}

Using \eqref{eq:thm4pe1} and \eqref{eq:thm4pe2} in \eqref{appBdef}, we find that the total error probability averaged over all possible modified sign-codes $\oPe \leq 2\varepsilon$ for $n$ large enough.
This implies the existence of a modified sign-code with total error probability $\Pe = \Pr\{(\hat{M}_a,\hat{M}_s)\neq (M_a,M_s)\} \leq 2\varepsilon$.
This holds for all $\varepsilon>0$, and~thus, the~rate:
\begin{equation}
 R = H(\boldB)+\gamma \leq \rbmd,
\end{equation}
is achievable with modified sign-coding, which concludes the proof of Theorem~\ref{mod_bmd}.

\bibliographystyle{IEEEtran}
\bibliography{IEEEabrv,MDPI_refs.bib}

\end{document}